\documentclass[aps,preprintnumbers,nofootinbib,noshowpacs,eqsecnum,prd,superscriptaddress,letterpaper]{revtex4} 

\usepackage[utf8]{inputenc}
\usepackage{graphicx}
\usepackage{amsmath,amssymb}
\usepackage{url}
\usepackage{multirow}
\usepackage{hyperref,color}
\usepackage[normalem]{ulem}
\graphicspath{{figs/}}
\newcommand{\sla}[1]{/\!\!\!\!#1}

\newcommand{\Dfb}{\mbox{$\raisebox{1mm}{\boldmath ${}^\leftrightarrow$} \hspace{-4mm} D$}}
\newcommand{\Dfba}{\mbox{$\raisebox{1mm}{\boldmath ${}^\leftrightarrow$}\hspace{-4mm} D^a$}}
\newcommand {\fbw}     {f_{BW}}
\newcommand {\fpone}   {f_{\Phi,1}}
\newcommand {\fwww}    {f_{WWW}}
\newcommand {\fw}      {f_{W}}
\newcommand {\fb}      {f_{B}}
\newcommand {\fww}     {f_{WW}}
\newcommand {\fbb}     {f_{BB}}
\newcommand {\fgg}     {f_{GG}}
\newcommand {\fptwo}   {f_{\Phi,2}}

\newcommand {\obw}     {{\cal O}_{BW}}
\newcommand {\opone}   {{\cal O}_{\Phi,1}}
\newcommand {\owww}    {{\cal O}_{WWW}}
\newcommand {\ow}      {{\cal O}_{W}}
\newcommand {\ob}      {{\cal O}_{B}}
\newcommand {\oww}     {{\cal O}_{WW}}
\newcommand {\obb}     {{\cal O}_{BB}}
\newcommand {\ogg}     {{\cal O}_{GG}}
\newcommand {\optwo}   {{\cal O}_{\Phi,2}}

\newcommand {\oqthree}  {{\cal O}^{(3)}_{\Phi,Q}}
\newcommand {\oqone}  {{\cal O}^{(1)}_{\Phi,Q}}
\newcommand {\our}  {{\cal O}^{(1)}_{\Phi,u}}
\newcommand {\odr}  {{\cal O}^{(1)}_{\Phi,d}}
\newcommand {\oud}  {{\cal O}^{(1)}_{\Phi,ud}}
\newcommand {\oer}  {{\cal O}^{(1)}_{\Phi,e}}
\newcommand {\ollll}  {{\cal O}_{LLLL}}
\newcommand {\ot}  {{\cal O}_{u\Phi,33}}
\newcommand {\otg}  {{\cal O}_{tG}}
\newcommand {\obo}  {{\cal O}_{d\Phi,33}}
\newcommand {\ota}  {{\cal O}_{e\Phi,33}}
\newcommand {\omu}  {{\cal O}_{e\Phi,22}}            
\newcommand {\fqthree}  {f^{(3)}_{\Phi,Q}}
\newcommand {\fqone}  {f^{(1)}_{\Phi,Q}}
\newcommand {\fur}  {f^{(1)}_{\Phi,u}}
\newcommand {\fdr}  {f^{(1)}_{\Phi,d}}
\newcommand {\fud}  {f^{(1)}_{\Phi,ud}}
\newcommand {\fer}  {f^{(1)}_{\Phi,e}}
\newcommand {\fllll}  {f_{LLLL}}
\newcommand {\fbo}  {f_{b}}
\newcommand {\fta}  {f_{\tau}}
\newcommand {\ft}  {f_{t}}
\newcommand {\ftG}  {f_{tG}}
\newcommand {\fm}  {f_{\mu}}
\preprint{\bf YITP-SB-2021-13}

\begin{document}

\title{Electroweak legacy of the LHC Run II}
\author{ Eduardo da Silva Almeida}
\email{eduardo.silva.almeida@usp.br}
\affiliation{Instituto de F\'isica, Universidade de S\~ao Paulo,
  R. do Mat\~ao 1371, 05508-090 S\~ao Paulo, Brazil}
\author{Alexandre Alves}
\email{aalves@unifesp.br}
\affiliation{Departamento de F\'isica, Universidade Federal de S\~ao Paulo, UNIFESP,
Diadema, S\~ao Paulo, Brazil}
\author{Oscar J. P. \'Eboli}
\email{eboli@if.usp.br}
\affiliation{Instituto de F\'isica, Universidade de S\~ao Paulo,
  R. do Mat\~ao 1371, 05508-090 S\~ao Paulo, Brazil}
\author{M.~C.~Gonzalez--Garcia}
\email{maria.gonzalez-garcia@stonybrook.edu}
\affiliation{Departament  de  Fisica  Quantica  i  Astrofisica
 and  Institut  de  Ciencies  del  Cosmos,  Universitat
 de Barcelona, Diagonal 647, E-08028 Barcelona, Spain}
\affiliation{Instituci\'o Catalana de Recerca i Estudis Avancats (ICREA)
Pg. Lluis  Companys  23,  08010 Barcelona, Spain.}
\affiliation{C.N. Yang Institute for Theoretical Physics, Stony Brook University, Stony Brook NY11794-3849,  USA}
%

%---------------------------------------------
\begin{abstract}
  We present a comprehensive study of the electroweak interactions
  using the available Higgs and electroweak diboson production results
  from LHC Runs 1 and 2 as well as the electroweak precision data in
  terms of the dimension-six operators.  Under the assumption that no
  new tree level sources of flavor violation nor violation of
  universality of the weak current is introduced, the analysis
  involves 21 operators.  We assess the impact of the data on
  kinematic distributions for the Higgs production at the LHC by
  comparing the results obtained including the simplified template
  cross section data with those in which only total Higgs signal
  strengths are considered.  We also compare the results obtained when
  including the dimension-six anomalous contributions to order
  $1/\Lambda^2$ and to order $1/\Lambda^4$.  As an illustration of the
  LHC potential to indirectly learn about specific forms of new
  physics, we adapt the analysis to constrain the parameter space for
  a few simple extensions of the standard model which generate a
  subset of the dimension-six operators at tree level.
\end{abstract}
%---------------------------------------------

\maketitle
%\tableofcontents
%\newpage

%%%%%%%%%%%%%%%%%%%%%%%%%%%%%%%%%%%%%%%%%%%%%%
\section{Introduction}
\label{sec:intro}

The large integrated luminosity accumulated by ATLAS and CMS during
the Run 2 of the Large Hadron Collider (LHC) extended the reach of
direct searches for new physics. It also allowed studying some
processes in more detail \emph{e.g.} differential distributions are
available for the Higgs production, or new channels were employed to
explore the self-interactions of the electroweak gauge bosons.  The
outcome of the direct searches has been negative so far: no new states
were observed. The most straight-forward interpretation of this result
points out to a higher-than-kinematically-accessible new physics
scale. This makes the framework of effective lagrangians
\cite{Weinberg:1978kz, Georgi:1985kw, Donoghue:1992dd} the obvious
tool to search for indirect signals of (or limits on) new physics.
\smallskip

The effective lagrangian approach is suited for model--independent
analyses since it is based exclusively on the low-energy accessible
states and symmetries.  Assuming that the scalar particle observed in
2012~\cite{ Aad:2012tfa, Chatrchyan:2012xdj} belongs to an electroweak
doublet, we can realize the $SU(2)_L \otimes U(1)_Y$ symmetry
linearly. The resulting model is the so called standard model
effective field theory (SMEFT).  In this framework, dimension-six
operators are the ones that can contribute to the LHC physics at the
lowest order. But even limiting the effective lagrangian to
dimension six the number of operators which can give signals at LHC is
large, stressing the relevance of the new kinematic information and
the new observed channels.  \smallskip

In this work, we focus on the part of the dimension-six lagrangian
that modifies the electroweak interactions. Under the assumption of no
new tree level sources of flavor violation or violation of
universality of the weak current, 21 dimension-six operators are
relevant, and we introduce in Sec.~\ref{sec:thframe} our choice of
operator basis.  Their Wilson coefficients parametrize our ignorance
of the specific form of the new physics and to obtain the strongest
possible bounds in this general scenario, one must perform global
analyses using all available experimental information. Here, we take
into account the electroweak precision data (EWPD) and the data on
electroweak diboson and Higgs productions at the LHC (see
Sec.~\ref{sec:thframe}).  Recently, ATLAS and CMS released information
on the kinematic distributions for the Higgs in the form of simplified
template cross sections
(STXS)~\cite{LHCHiggsCrossSectionWorkingGroup:2016ypw,
  Andersen:2016qtm}. In order to understand the impact of such
distributions with the present integrated luminosity, we have
performed our studies using the STXS data as well as the Higgs results
in the form of total signal strength (SS).  \smallskip

In a bottom-up approach all Wilson coefficients are treated as free
parameters. But, in a realistic UV completion, the Wilson coefficients
are determined by the high energy physics and might be correlated. In
order to illustrate the importance of these correlations, we also
obtain constraints on simple extensions of the standard model (SM)
that give rise to dimension-six operators at tree level. We consider
extensions with only one single particle added to the spectrum and two
Higgs doublet models which we introduce in
Sec.~\ref{sec:models}. \smallskip

The results stemming from our global analyses, presented in
Sec.~\ref{sec:results} and Sec.~\ref{sec:discussion}, show no
statistically significant source of tension with the SM.  We stress
the complementarity of the different data sets in reaching this
conclusion.  In particular we find that the LHC integrated luminosity
is large enough to allow for non-negligible tightening of the bounds
on some of the Wilson coefficients already constrained by the
EWPD. Moreover, we show that the use of Higgs data is important to
limit departures from the SM predictions for the triple electroweak
gauge couplings (TGC) due to the correlation introduced by the linear
realization of the SM gauge symmetry in the effective
operators. \smallskip

Our analyses were conducted using the theoretical predictions linear
(i.e. ${\cal O}(\Lambda^{-2})$) in the Wilson coefficients, as well as
including also the quadratic (${\cal O}(\Lambda^{-4})$) terms with the
aim at estimating the uncertainties associated to the expansion in
power of the high scale $\Lambda$. When working up to order
$\Lambda^{-4}$ a set of (quasi)degenerate solutions appear associated
to the flip of the sign of some Higgs coupling. They lead to
 disconnected allowed ranges for the corresponding Wilson coefficients
which do not contain the SM.  Our analysis show how Higgs kinematic
distributions, do not only reduce the allowed SM-connected solutions,
but they also help in resolving some of these degeneracies ruling out
the non-SM-connected solutions for some of the operators. Generically
we find that, when focusing on SM-connected ranges of parameters, the
precision on the bounds derived at ${\cal O}(\Lambda^{-2})$ and at
${\cal O}(\Lambda^{-4})$ are becoming comparable for most operators.
Quantitatively our results are summarized in Table ~\ref{tab:ranges}
and Figs.~\ref{fig:franges} and ~\ref{fig:scales}. \smallskip

At present the comprehensive analysis of collider results in the
framework of the SMEFT is in the hands of phenomenologists (see
Refs.~\cite{Falkowski:2019hvp, Dawson:2020oco, DeBlas:2019ehy,
  Anisha:2020ggj, Ethier:2021bye, Ellis:2020unq} for most recent
analysis). This article adds to this literature by including the most
updated data-sets and by the different working hypothesis employed.
In brief, some of those analysis make a detailed study of flavor
effects, however, they do not take into account the Higgs kinematic
distribution in the STXS format; see for
instance~\cite{Falkowski:2019hvp, DeBlas:2019ehy, Anisha:2020ggj}.
From the point of view of data samples included in the analysis,
Refs.~\cite{Ellis:2020unq,Ethier:2021bye} are the most complete also
including top results. But in what respects Higgs analysis they did
not include the full updated ATLAS and CMS STXS of Run 2 Higgs data
which we consider here. This allows us to quantify the present impact
of the Higgs kinematic distribution data by contrasting the bounds
obtained using STXS with that using only the signal strength data. In
what respect to the working hypothesis, we perform our analyses using
the HISZ basis ~\cite{Hagiwara:1993ck, Hagiwara:1996kf} that does not
exhibit blind directions in the EWPD in contrast with the Warsaw
one~\cite{Grzadkowski:2010es} employed in
Refs.~\cite{Falkowski:2019hvp, Dawson:2020oco, DeBlas:2019ehy,
  Anisha:2020ggj, Ethier:2021bye, Ellis:2020unq}; see
section~\ref{sec:thframe}.  Finally , as mentioned above, we perform
our analysis both at order ${\cal O}(\Lambda^{-2})$, and at
${\cal O}(\Lambda^{-4})$ as a way to quantitatively address the
importance of higher order terms and the use of EFT within its range
of validity.  \smallskip

%%%%%%%%%%%%%%%%%%%%%%%%%%%%%%%%%%%%%%%%%%%%%%%%%%%%%%%%%%%%%%%%%%%%%%
\section{Theoretical Framework}
\label{sec:thframe}

In order to describe the effects of new physics only directly
accessible at higher energy scale $\Lambda$, we supplement the SM with
higher-dimension operators
\begin{equation}
   {\cal L}_{\rm eff} = {\cal L}_{\rm SM} + \sum_{n>4,j}
   \frac{f_{n,j}}{\Lambda^{n-4}} {\cal O}_{n,j} \;,
\label{eq:gen}
\end{equation}
where the SM $SU(3)_C \otimes SU(2)_L \otimes U(1)_Y$ gauge symmetry
is realized linearly in the ${\cal O}_{n,j}$ operators. Neutrino
physics strongly constrains the dimension-five
operator~\cite{Weinberg:1979sa}, therefore dimension-six operators are
those with lowest dimension that contribute to processes at the
LHC.  For the sake of simplicity we focus on dimension-six operators
which conserve $C$ and $P$ (besides total lepton and baryon number).
It is well known that there are 59 independent dimension-six
operators~\cite{Grzadkowski:2010es}, up to flavor and hermitian
conjugation.  Here, we focus on those that modify the electroweak
interactions impacting precision electroweak data, TGC's and Higgs
physics.  For our analysis 21 independent dimension-six operators are
relevant and we list our choice of basis in Tables~\ref{tab:operbos}
and ~\ref{tab:operfer}.  \smallskip

In particular, for the pure bosonic operators we use the Hagiwara,
Ishihara, Szalapski, and Zeppenfeld (HISZ) basis
~\cite{Hagiwara:1993ck, Hagiwara:1996kf} (see
Table.~\ref{tab:operbos}).  $\Phi$ stands for the SM Higgs doublet and
we have defined $\widehat{B}_{\mu\nu} \equiv i(g^\prime/2)B_{\mu\nu}$
and $\widehat{W}_{\mu\nu} \equiv i(g/2)\sigma^aW^a_{\mu\nu}$, with $g$
and $g^\prime$ being the $SU(2)_L$ and $U(1)_Y$ gauge couplings,
respectively.  In what respects the main effect of these
operators in the observables, the $\obw$ and $\opone$ operators are
ubiquitous: they modify the electroweak gauge boson couplings among
themselves and to the Higgs boson and fermions. The operator $\optwo$
rescales all the SM Higgs couplings. $\ob$, $\ow$ and $\owww$
contribute to triple gauge couplings, while $\ob$, $\ow$, $\obb$,
$\oww$, and $\ogg$ affect the Higgs couplings to gauge
bosons. \smallskip

We notice that our choice of operators is at difference with the
independent basis choice in Ref.~\cite{Grzadkowski:2010es} and which
is employed in other analysis (see Refs.~\cite{Ellis:2014jta,
  Falkowski:2014tna, Ellis:2020unq, Ellis:2018gqa, Dawson:2020oco,
  Falkowski:2019hvp, DeBlas:2019ehy, Anisha:2020ggj} for some recent
examples and references therein).  Here, we keep the operators
${\cal O}_B$ and ${\cal O}_W$ and we remove instead some operators
involving fermions. \smallskip

%%%%%%%%%%%%%%%%%%%%%%%%%%%%%%%%%%%%%%%%%%%%%%%%%%%%%%%
\begin{table}[h]
  \begin{tabular}{|l|l|l|}
    \hline
    $\obw = \Phi^\dagger\widehat{B}_{\mu\nu}\widehat{W}^{\mu\nu}\Phi$
    & $\opone =(D_\mu\Phi)^\dagger\Phi\Phi^\dagger(D^\mu\Phi)  $
    &  $\optwo =\frac{1}{2} \partial^\mu\left ( \Phi^\dagger \Phi \right)
                            \partial_\mu\left ( \Phi^\dagger \Phi
  \right )$
                        \\
    \hline
    $\ob =	(D_\mu\Phi)^\dagger\widehat{B}^{\mu\nu}(D_\nu\Phi) $
    & $\ow =	(D_\mu\Phi)^\dagger\widehat{W}^{\mu\nu}(D_\nu\Phi) $
    & $\owww=  {\rm Tr}[\widehat{W}_{\mu}^{\nu}
    \widehat{W}_{\nu}^{\rho}\widehat{W}_{\rho}^{\mu}]$
    \\
    \hline
    $\ogg  = \Phi^\dagger \Phi \; G_{\mu\nu}^a G^{a\mu\nu}  $
    & $\obb = \Phi^{\dagger} \hat{B}_{\mu \nu} \hat{B}^{\mu \nu} \Phi $
    & $\oww  = \Phi^{\dagger} \hat{W}_{\mu \nu} \hat{W}^{\mu \nu} \Phi $
    \\
    \hline
  \end{tabular}
  \caption{Dimension-six operators containing only bosonic fields that we
    considered in our analyses. The notation is given in the text.}
  \label{tab:operbos}
\end{table}
%%%%%%%%%%%%%%%%%%%%%%%%%%%%%%%%%%%%%%%%%%%%%%%%%%%%%%%
%-------------------------------------------------------
\begin{table}[h]
  \begin{tabular}{|l|l|l|}
    \hline
     $ {\cal O}^{(1)}_{\Phi Q}=
    \Phi^\dagger (i\,\Dfb_\mu \Phi)  
    (\bar Q\gamma^\mu Q) $
    &
    ${\cal O}^{(3)}_{\Phi Q} =\Phi^\dagger (i\,{\Dfba}_{\!\!\mu} \Phi) 
  (\bar Q\gamma^\mu T_a Q)$
    &
      ${\cal O}^{(1)}_{\Phi u}=\Phi^\dagger (i\,\Dfb_\mu \Phi) 
      (\bar u_{R}\gamma^\mu u_{R})$
    \\
    \hline
      $ {\cal O}^{(1)}_{\Phi d}=\Phi^\dagger (i\,\Dfb_\mu \Phi) 
      (\bar d_{R}\gamma^\mu d_{R})$
        &
    ${\cal O}^{(1)}_{\Phi ud}=\tilde\Phi^\dagger (i\,\Dfb_\mu \Phi) 
    (\bar u_{R}\gamma^\mu d_{R})$
    &     ${\cal O}^{(1)}_{\Phi e}=\Phi^\dagger (i\,\Dfb_\mu \Phi) 
    (\bar e_{R}\gamma^\mu e_{R})$
    \\
    \hline
    ${\cal O}_{tG} = (\bar{Q_3}\sigma^{\mu\nu} \frac{\lambda^a}{2} {u_3}) \tilde{\Phi}G^a_{\mu\nu}$
    &
    &
      \\
    \hline\hline
    ${\cal O}_{e\Phi,ii}=(\Phi^\dagger\Phi)(\bar L_i \Phi e_{R,i}) $ \; $i=2,3$
    & ${\cal O}_{u\Phi,33}=(\Phi^\dagger\Phi)(\bar Q_3 \tilde \Phi u_{R,3})$
    &${\cal O}_{d\Phi,33}=(\Phi^\dagger\Phi)(\bar Q_3 \Phi d_{R,3})$
    \\
    \hline\hline
       ${\cal O}_{LLLL}
      =(\bar L \gamma^\mu L)(\bar L \gamma^\mu L)$
    &
    &
    \\
    \hline
  \end{tabular}
  \caption{Dimension-six operators containing fermionic fields that we
    considered in our analyses. The notation is given in the text.}
  \label{tab:operfer}
\end{table}
%------------------------------------------------------

The independent set of fermionic operators employed in our analysis is
presented in Table~\ref{tab:operfer}, where the lepton (quark) doublet
is denoted by $L$ ($Q$) and $f_R$ are the $SU(2)_L$ singlet
fermions. When including a sub-index it refers to the fermionic
generation. When no sub-index is included the operator is the same for
three generations. In addition, we defined
$\Phi^\dagger\Dfb_\mu\Phi= \Phi^\dagger D_\mu\Phi-(D_\mu\Phi)^\dagger
\Phi$ and
$\Phi^\dagger \Dfba_{\!\!\mu} \Phi= \Phi^\dagger T^a D_\mu
\Phi-(D_\mu\Phi)^\dagger T^a \Phi$ with $T^a=\sigma^a/2$ with
$\sigma^a$ being the Pauli matrices. The Gell-Mann matrices are
represented by $\lambda^a$.\footnote{We did not considered six dipole
  operators whose interference with the SM contributions vanish for
  EWPD observables. As shown in Ref.~\cite{daSilvaAlmeida:2019cbr}
  including those additional operators would have no impact on the
  determination of the operators considered here.}  In selecting the
operators in Table~\ref{tab:operfer} we have assumed no family mixing
to prevent the generation of too large flavor violation
~\cite{Alonso:2013hga, Henning:2015alf} and, for the sake of
simplicity, we have considered the operators involving couplings to
gauge bosons to be generation independent.  With these hypotheses, the
operators involving lepton doublets ${\cal O}^{(1)}_{\Phi L}$ and
${\cal O}^{(3)}_{\Phi L}$ are removed from our
basis~\cite{Corbett:2012ja} employing the freedom associated to the
use of equations of motion~\cite{Politzer:1980me, Georgi:1991ch,
  Arzt:1993gz, Simma:1993ky}. This, as mentioned above, is at
difference with the choice of basis in Ref.~\cite{Grzadkowski:2010es}
in which these lepton doublet operators are kept in exchange of the
bosonic operators ${\cal O}_B$ and ${\cal O}_W$. The advantage of our
choice is that it avoids the existence of blind
directions~\cite{DeRujula:1991ufe,Elias-Miro:2013mua} in the EWPD
analyses so that the results of EWPD can be shown independently. It
also allows for better discrimination of the impact of the LHC data on
the precise determination of the fermion-gauge couplings versus gauge
boson self-couplings.  \smallskip

In summary, the operators in the first two lines of
Table~\ref{tab:operfer} modify the fermion couplings to electroweak
gauge bosons and the Higgs.  And we notice that with our choice of
basis the only operator that modifies the $Z$ coupling to leptons is
${\cal O}^{(1)}_{\Phi e}$.  On the other hand, the operator in the
third line is the only one in our analyses affecting the fermion-gluon
coupling and it was included since it impacts the Higgs production via
gluon fusion. The operators modifying the Yukawa interactions are
presented in the fourth line, while the last one contains the only
four-fermion operator that takes part in the determination of the
Fermi constant.\smallskip

Altogether eight operators contribute at ${\cal O}(\Lambda^{-2})$ at
tree level to the EWPD observables~\cite{ALEPH:2005ab} and the
relevant part of the effective lagrangian reads
\begin{eqnarray}
 \Delta  {\cal L}_{\rm eff}^{\rm EWPD}
  &=& \frac{f^{(1)}_{\Phi Q}}{\Lambda^2}   {\cal O}^{(1)}_{\Phi Q}
  +     \frac{f^{(3)}_{\Phi Q}}{\Lambda^2}   {\cal O}^{(3)}_{\Phi Q}
  +     \frac{f^{(1)}_{\Phi u}}{\Lambda^2}   {\cal O}^{(1)}_{\Phi u}
  +     \frac{f^{(1)}_{\Phi d}}{\Lambda^2}   {\cal O}^{(1)}_{\Phi d}
%  +     \frac{f^{(1)}_{\Phi ud}}{\Lambda^2}   {\cal O}^{(1)}_{\Phi ud} 
  +     \frac{f^{(1)}_{\Phi e}}{\Lambda^2}   {\cal O}^{(1)}_{\Phi e}
\nonumber
\\
&+&
\frac{f_{BW}}{\Lambda^2} {\cal O}_{BW}
+ \frac{f_{\Phi,1}}{\Lambda^2} {\cal O}_{\Phi,1}
  + 2    \frac{f_{LLLL}}{\Lambda^2}   {\cal O}_{LLLL} \;\;. 
\label{eq:leff-ewpd}
\end{eqnarray}

The production of electroweak gauge boson pairs $W^+ W^-$, $ W^\pm Z$
and $W^\pm \gamma$, (here on EWDBD), receives contributions from
dimension-six operators that modify the electroweak gauge boson
couplings to fermions, as well as the triple gauge couplings. It is
interesting to notice that dimension–-six operators do not give rise
to anomalous TGC among neutral gauge bosons, which appear only at
dimension eight~\cite{Degrande:2013kka}. The part of the effective
lagrangian that contributes to EWDBD, in addition to that
participating in the EWPD analysis, is
\begin{equation}
\Delta{\cal L}_{\rm eff}^{\rm TGC} = 
 \frac{f_{WWW}}{\Lambda^2} {\cal  O}_{WWW}
+ \frac{f_{W}}{\Lambda^2} {\cal O}_{W}
+ \frac{f_{B}}{\Lambda^2} {\cal O}_{B}
+ \frac{f^{(1)}_{\Phi ud}}{\Lambda^2} \left(  {\cal O}^{(1)}_{\Phi
ud} + {\rm h.c.} \right) \;\;.
\label{eq:leff-tgc}
\end{equation}
The last operator modifies just the right-handed $W$ couplings to
quarks, and consequently, it does not contribute to the EWPD
observables at ${\cal O}(\Lambda^{-2})$.\smallskip

Dimension-six operators modify the Higgs production and decay through
changes in its couplings to fermions and gauge bosons.  We parametrize
the contributions exclusive to the Higgs physics as
\begin{eqnarray}
\Delta   {\cal L}_{\rm eff}^{\rm H} &=& 
\frac{f_\mu m_\mu}{\Lambda^2 v} {\cal O}_{e\Phi,22} 
+ \frac{f_\tau m_\tau}{ \Lambda^2 v} {\cal O}_{e\Phi,33} 
+ \frac{f_b m_b}{ \Lambda^2 v} {\cal O}_{d\Phi,33} 
+ \frac{f_t m_t}{ \Lambda^2 v} {\cal O}_{u\Phi,33} 
+ \text{ h.c.}
  \nonumber
  \\
&-& \frac{\alpha_s }{8 \pi} \frac{f_{GG}}{\Lambda^2} {\cal O}_{GG}  
+ \frac{f_{BB}}{\Lambda^2} {\cal O}_{BB} 
+ \frac{f_{WW}}{\Lambda^2} {\cal O}_{WW} 
    + \frac{f_{\Phi,2}}{\Lambda^2} {\cal O}_{\Phi,2}
    + \frac{f_{tG}}{\Lambda^2} {\cal O}_{tG}\;\;.
   \label{eq:leff-h}
\end{eqnarray}

In summary, we considered the total effective lagrangian 
\begin{equation}
{\cal L}_{\rm eff} = {\cal L}_{\rm SM} 
+\Delta {\cal L}_{\rm eff}^{\rm EWPD}
+ \Delta {\cal L}_{\rm eff}^{\rm TGC}
+ \Delta {\cal L}_{\rm eff}^{\rm H} \;\;.
\label{eq:leff}
\end{equation}
For a complete list of vertices generated by Eq.~(\ref{eq:leff}), see
reference~\cite{daSilvaAlmeida:2018iqo}.  The explicit form of the
couplings and the different Lorentz structures generated can be found,
for example, in Refs.~\cite{Corbett:2012dm, Corbett:2012ja,
  Corbett:2017qgl, Corbett:2014ora, Alves:2018nof} to which we refer
the reader for details. \smallskip

As we will see in the following section, at present there is enough
experimental information to individually bound the 21 Wilson
coefficients.  But it is important to notice that some Wilson
coefficients can lead to a sign change of Higgs couplings with respect
to the SM predictions, consequently leading to discrete (quasi)
degeneracies in the global analyses. For instance, the modification of
the coefficient of the $H W^+_\mu W^{- \mu}$ vertex is
\begin{equation}
\left ( \frac{g^2 v}{2} \right ) 
\left [
1 - \frac{v^2}{4} 
\left (
 \frac{f_{\Phi, 1}}{\Lambda^2}+ 2 \frac{f_{\Phi,2}}{\Lambda^2}
\right ) 
\right ] \;\;.
\label{eq:vert-hww} 
\end{equation}
Since $f_{\Phi,1}/\Lambda^2$ is strongly bounded by EWPD, it is only
possible to have a degeneracy with the SM results for both
$f_{\Phi,2}/\Lambda^2=0$ and
$f_{\Phi,2}/\Lambda^2\sim 4 / v^2 \sim 65$ TeV$^{-2}$.  These points
in parameter space are also nearly degenerate for the vertex
$H Z_\mu Z^\mu$. \smallskip

In similar fashion the anomalous interactions can also lead to Yukawa
couplings of the order of the SM ones but with a different sign as the
coefficient of the $H \bar{f} f$ vertex is now
\begin{equation}
- \frac{m_f}{v} \left [
1 - \frac{v^2}{4} 
\left (\frac{f_{\Phi,1}}{\Lambda^2}+
2\frac{f_{\Phi,2}}{\Lambda^2}
+ 2\,\sqrt{2} \frac{f_f}{\Lambda^2} 
\right )
\right ]
\label{eq:vert-yuk}
\end{equation}
where $f= \mu, \tau, b, t$. Therefore, $f_{\Phi,2}/\Lambda^2$ and
$f_f/\Lambda^2$ can flip the sign of the Yukawa coupling.  In
combination with the 2-fold degeneracy $f_{\Phi,2}/\Lambda^2$ in
Eq.~\eqref{eq:vert-hww} there are 2x2 quasi-degenerate SM-like
solutions: two for $f_f/\Lambda^2= 0$ and two with
$f_f/\Lambda^2\pm 2\sim \sqrt{2}/v^2\sim 45$
TeV$^{-2}$~\cite{daSilvaAlmeida:2018iqo}.  \smallskip

Another potential source of approximate degeneracies/correlations is
the effective photon-photon-Higgs coupling $H F_{\mu\nu} F^{\mu\nu}$
for which the dimension-six operators induce a tree-level contribution
\begin{equation} 
  A(\gamma\gamma\to H) = A_{SM}(\gamma\gamma\to H) +
  \frac{e^2 v}{4} \frac{f_{WW}+f_{BB}-f_{BW}}{\Lambda^2} \;\;,
  \label{eq:vert-gaga}
\end{equation}
where $A_{SM}(\gamma\gamma\to H)\simeq { -8.25\times 10^{-3}}$
TeV$^{-1}$. First, taking into account the strong EWPD bounds on
$f_{BW}$ this dependence leads to a strong anti-correlation between
the allowed values of two  remaining Wilson coefficients.
That correlation is partly broken by the measurement of the effective
photon-Z-coupling $H F_{\mu\nu} Z^{\mu\nu}$ which constrains a
different combination of $f_{WW}/\Lambda^2$, $f_{BB}/\Lambda^2$ and
$f_{BW}/\Lambda^2$. Furthermore it is possible to find a SM-like
solutions for $(f_{WW}+f_{BB})/\Lambda^2 \sim 3$ TeV$^{-2}$ with
inverted sign of the photon-photon-Higgs effective coupling. \smallskip

The equivalent effect is also present in the gluon-gluon-Higgs
interaction.  In the large top mass limit the lowest order amplitude
is \footnote{Notice that we have defined the Wilson coefficient of the
  gluon-gluon-Higgs operator $f_{GG}$ in Eq.(\ref{eq:leff-h}) to
  include a loop-like suppression factor so even if it gives a
  tree-level contribution to the amplitude it can be factorized in
  this form in the large top mass limit. We work here in a different
  convention than Refs.~\cite{Corbett:2012ja,daSilvaAlmeida:2018iqo}
  and do not include in the definition of $ {\cal O}_{GG}$ the effects
  of the anomalous operators that modify the couplings of the top
  quarks running in the loop.}
\begin{equation}
A(gg\to H) = A_{SM}(gg\to H) \left[
  1+ \frac{3}{2} v^2 \frac{\fgg}{\Lambda^2}
  -\frac{v^2}{4} \left( \frac{\fpone}{\Lambda^2} + 2
  \frac{\fptwo}{\Lambda^2} + 2 \sqrt{2} \frac{\ft}{\Lambda^2} \right)
  - \frac{\sqrt{2\pi\alpha_s} v m_t^3}{m_H^2} \frac{\ftG}{\Lambda^2}
\right ] \;\;.
\label{eq:vert-gluglu}
\end{equation}

It is important to notice that in most cases the operators
contributing to the Higgs vertices in
Eqs.~\eqref{eq:vert-hww}--\eqref{eq:vert-gluglu} involve different
number of derivatives of the Higgs field. Consequently, we can
anticipate the relevance of the data on kinematic distributions to
resolve the vertex degeneracies and limit the
correlations~\cite{Grojean:2013nya, Buschmann:2014twa,
  Dawson:2014ora}, in particular for gluon-gluon-Higgs which
contributes to the main Higgs production mechanism.  \smallskip

%%%%%%%%%%%%%%%%%%%%%%%%%%%%%%%%%%%%%%%%%%%%%%%%%%%%%%%%%%%%%%%%%%%%%%
\subsection{Effective Lagrangians for Simplified Models}
\label{sec:models}

Effective lagrangians provide the framework for a bottom-up approach
for the search for new physics through deviations of the SM
predictions in a model independent fashion. The disadvantage of this
approach is that specific UV completions give rise, in general, to
relations between the Wilson coefficients of different operators,
which result into stronger constraints.  To illustrate the potential
of the effective lagrangians framework, we considered some simple
models whose matching at high energies are at tree level.  For the
sake of simplicity we do not consider the running of the Wilson
coefficients to the electroweak scale~\cite{Jenkins:2013zja,
  Jenkins:2013wua,Alonso:2013hga}. This renders the results
independent of the matching scale and can lead to bounds which can
either be slightly weaker or slightly stronger depending on the model
~\cite{Dawson:2020oco}.  \smallskip
\begin{itemize}
\item Extensions of the SM Higgs sector.  The simplest model in this
  category is obtained by adding a singlet scalar ($S$) which mixes
  with the SM Higgs with a mixing angle $\theta$.  An interesting
  feature of the most general form of this model is that it can
  exhibit a strong first-order electroweak phase transition that can
  be addressed in complementary studies by colliders and gravitational
  wave experiments~\cite{Profumo:2014opa, Chen:2017qcz}.  In here,
  however, we will focus on the case with a $Z_2$
  symmetry~\cite{Dawson:2020oco, Brivio:2021alv}.  At the high scale
  $\Lambda = M_s$ this model generates just one non-vanishing Wilson
  coefficient at tree level:
\begin{equation}
 \frac{f_{\Phi,2}}{\Lambda^2} = \frac{\tan^2\theta}{v^2} \;\;.
\label{eq:mod-sing}
\end{equation}
We also considered four classes of two Higgs doublet models (2HDM)
that satisfy the Weinberg-Glashow condition to avoid flavor changing
neutral currents~\cite{PhysRevD.15.1958}, and labeled as 2HDM Type-I,
2HDM Type-II, 2HDM lepton-specific and 2HDM flipped (for further
details see~\cite{Branco:2011iw, Gorbahn:2015gxa, Dawson:2020oco}).
The Wilson coefficients generated at tree level by these 2HDM in the
alignment limit ($|\cos(\beta-\alpha)| << 1$)
are~\cite{Gorbahn:2015gxa, Belusca-Maito:2016dqe,Dawson:2020oco}
\begin{equation}
  \frac{f_b}{\Lambda^2} = - \eta_b  \frac{\cos(\beta-\alpha)}{\tan\beta} \frac{\sqrt{2}}{v^2}
\;\;\;;\;\;\;
  \frac{f_t}{\Lambda^2} = - \eta_t  \frac{\cos(\beta-\alpha)}{\tan\beta} \frac{\sqrt{2}}{v^2}
\;\;\;;\;\;\;
  \frac{f_\tau}{\Lambda^2} = - \eta_\tau
  \frac{\cos(\beta-\alpha)}{\tan\beta} \frac{\sqrt{2}}{v^2}
  \;\;,
\label{eq:mod-2hdm}
\end{equation}
where $\tan\beta$ is the ratio between the Higgs doublet vevs and
$\alpha$ is the neutral Higgs mixing angle. The $\eta$ factor depends
on the 2HDM type:
$$  \begin{array}{ l|c|c|c}
    {\rm 2HDM}& \eta_t & \eta_b & \eta_\tau
    \\
    \hline
    {\rm Type-I} & 1 & 1 &1
    \\
    {\rm Type-II} &  1& -\tan^2\beta & -\tan^2\beta
    \\
    {\rm Lepton-specific} & 1 & 1 & -\tan^2\beta
    \\
    {\rm Flipped} & 1 & -\tan^2\beta &  1
    \\
  \end{array}
$$

\item Simplified models~\cite{LHCNewPhysicsWorkingGroup:2011mji} that
  contain the addition of one new (no-scalar) state.  We consider five
  different charge assignments for the additional state as listed in
  Table~\ref{tab:models} (see Ref.\cite{deBlas:2017xtg}).  We also
  list in the third line the SM particles which couple to the new
  state. For consistency with our EFT assumptions, we assume equal
  couplings of the new state to the three families. And for simplicity
  we also assume that the nonzero couplings of the new state to the
  SM particles have the same strength $\chi$.
%%%%%%%%%%%%%%%%%%%%%%%%%%%%%%%%%%%%%%%%%%%%%%%%%%%%%%%
\begin{table}[h]
  \begin{tabular}{|c|c|c|c|c|c|}
      \hline
      Additional state & Vector $B$& Vector $B'$ &Vector $W$ & Lepton $E$ & Quark $U$
\\\hline
Rep $(SU(3)_c \;,\; SU(2)_L\;,\; U(1)_Y)$ &$(1 \;,\; 1 \;,\; 0)$
& $(1 \;,\; 1 \;,\; 0)$ & $(1 \;,\; 3 \;,\; 0)$
& $(1 \;,\; 1 \;,\; -1)$ & $(3 \;,\; 1 \;,\; 2/3)$
    \\\hline
    Couplings
    & 
    Higgs
    &  Higgs + quarks
    &  Higgs + quarks
    &  leptons
    &  quarks
    \\\hline\hline
    $\fpone/\Lambda^2$ & -2 & -2 & -- & -- & --
    \\
    \hline
    $\fptwo/\Lambda^2$ &   1 &  1 & 3/4 & -- & --
    \\
    \hline
    $\fqone/\Lambda^2$ & -- & -1 & -- & -- & 1/4
    \\
    \hline
    $\fqthree/\Lambda^2$ & -- & -- & -1/4 & -- & -1/4
    \\
    \hline 
    $f^{(1)}_{\Phi u}/\Lambda^2$ & -- & -1 & -- & -- & --
    \\
    \hline
    $f^{(1)}_{\Phi d}/\Lambda^2$ & -- & -1 & -- & -- & --
    \\
    \hline
    $\fbo/\Lambda^2$ & -- & -- & -$\sqrt{2}$/4
    & -- & --
    \\
    \hline
    $\ft/\Lambda^2$ & -- & -- & -$\sqrt{2}$/4 & -- & $\sqrt{2}$/2\,
    \\
    \hline
    $\fta/\Lambda^2$ & -- & -- & -- & $\sqrt{2}$/2\, & --
\\
    \hline
    $\fm/\Lambda^2$ & -- & -- & -- & $\sqrt{2}$/2\, & --
    \\
    \hline 
    $f^{(1)}_{\Phi \ell}/\Lambda^2$ & -- &-- & -- & -1/4 & --
    \\
    \hline
    $f^{(3)}_{\Phi \ell}/\Lambda^2$ & -- &-- & -- & -1/4 & --
    \\
    \hline
  \end{tabular}
  \caption{New particle content and Wilson coefficients generated by
    extending the SM with such extra state. All Wilson coefficients
    must be multiplied by $\chi^2/M^2$, where $\chi$ parametrizes the
    universal strength of the allowed couplings of the new state to
    the SM particles and $M$ is its mass.}
  \label{tab:models}
\end{table}
%%%%%%%%%%%%%%%%%%%%%%%%%%%%%%%%%%%%%%%%%%%%%%%%%%%%%%%

With these assumptions the Wilson coefficients generated at tree level
by the new states are those listed in Table~\ref{tab:models}. Notice
that the existence of a heavy lepton $E(1,1)_{-1}$ generates two
operators that were removed from our basis. Consequently we have to
rotate the result back to our basis using that
\begin{eqnarray}
  {\cal O}^{(1)}_{\Phi \ell;jj} = && \frac{8}{g^{\prime\,2}} {\cal  O}_B -2 {\cal O}_{\Phi,2} 
                  +4   {\cal O}_{\Phi,1} +  \frac{4}{g^{\prime\,2}}   {\cal  O}_{BB} 
                   +  \frac{4}{g^{\prime\,2}}   {\cal  O}_{BW} 
\nonumber
  \\
  && -2  {\cal O}_{\Phi e;jj}  +\frac{1}{3}  {\cal O}^{(1)}_{\Phi
     Q;jj}  +\frac{4}{3}  {\cal O}_{\Phi u;jj}  -\frac{2}{3}  {\cal
     O}_{\Phi d;jj}  \;\;,
     \label{eq:rot1}
  \\
    {\cal O}^{(3)}_{\Phi \ell;jj} = && -\frac{8}{g^{2}} {\cal  O}_W + 6 {\cal O}_{\Phi,2} 
                  - \frac{4}{g^{2}}   {\cal  O}_{WW} 
                   -  \frac{4}{g^{2}}   {\cal  O}_{BW} 
\nonumber
  \\
  && -  {\cal O}^{(3)}_{\Phi Q;jj}  -2 \left[  Y^\dagger_u {\cal O}_{u
     \Phi}  + Y^\dagger_d {\cal O}_{d \Phi}  + Y^\dagger_e {\cal O}_{e
     \Phi}  
     \right] \;\;,
     \label{eq:rot2}
\end{eqnarray}
where $Y_f$ stands for the Yukawa coupling of the fermion $f$.

\end{itemize}

%%%%%%%%%%%%%%%%%%%%%%%%%%%%%%%%%%%%%%%%%%%%%%%%%%%%%%%%%%%%%%%%%%%%%%
\section{ANALYSES FRAMEWORK}
\label{sec:frame}

In order to study the Wilson coefficients of the dimension-six
operators of the effective lagrangian Eq.~(\ref{eq:leff}), we take
into account the EWPD, diboson production, and Higgs data on signal
strength and simplified template cross section measurements.  We
perform the global analyses either using only the contributions that
are linear on the Wilson coefficients (${\cal O}(1/\Lambda^2)$) or
including the quadratic terms (${\cal O}(1/\Lambda^4)$) as well.
\smallskip

In the EWPD analysis, we study 15 observables of which 12 are $Z$
observables~\cite{ALEPH:2005ab}:
\begin{equation}
\Gamma_Z \;\;,\;\;
\sigma_{h}^{0} \;\;,\;\;
{\cal A}_{\ell}(\tau^{\rm pol}) \;\;,\;\;
R^0_\ell \;\;,\;\;
{\cal A}_{\ell}({\rm SLD}) \;\;,\;\;
A_{\rm FB}^{0,l} \;\;,\;\;
R^0_c \;\;,\;\;
 R^0_b \;\;,\;\;
{\cal  A}_{c} \;\;,\;\;
 {\cal A}_{b} \;\;,\;\;
A_{\rm FB}^{0,c}\;\;,\;\;
\hbox{ and} \;\;
A_{\rm FB}^{0,b}  \hbox{ (SLD/LEP-I)}\;\;\; ,
\end{equation}
supplemented by three $W$ observables: the $W$ mass ($M_W$) taken
from~\cite{Olive:2016xmw}, its width ($\Gamma_W$) from
LEP2/Tevatron~\cite{ALEPH:2010aa} and the leptonic $W$ branching ratio
($\hbox{Br}( W\to {\ell\nu})$)~\cite{Olive:2016xmw}.\smallskip

The first component of our global $\chi^2$ function  is therefore suited for
the EWPD data
\begin{equation}
\chi^2_{\rm EWPD}(\fbw,\fpone,\fqthree,\fqone,\fur,\fdr,\fer,\fllll) \;.
\label{eq:chi2ewpd}
\end{equation}
We included in the EWPD analysis, the correlations among the above
observables, as displayed in Ref.~\cite{ALEPH:2005ab}, while the SM
predictions and uncertainties were extracted
from~\cite{Ciuchini:2014dea}.  For further details of this part of the
statistical analysis we refer the reader to
Refs.~\cite{Corbett:2017qgl, Alves:2018nof}.
\smallskip

%--Diboson data:

Triple gauge boson couplings have already been studied in the
production of electroweak gauge boson pairs at LEP2~\cite{lep2}, and the
Runs 1 and 2 of LHC~\cite{Aad:2016wpd, Khachatryan:2015sga,
  Aad:2016ett, Khachatryan:2016poo, ATLAS:2016qzn, ATLAS:2018ogj,
  ATLAS:2017pbb, CMS:2021lix, CMS:2020mxy, CMS:2021rym,
  Aaboud:2017gsl, ATLAS:2020nzk}. Here we analyze the most complete samples of
$W^+W^-$, $W^\pm Z$, $W^\pm \gamma$, and $Zjj$.
More specifically, the channels that we study and
the kinematic distribution included in the analysis  are listed
in Table~\ref{tab:diboson}. Further details on the analysis of EWDBD
from Run 1 can be seen at the Ref.~\cite{Alves:2018nof}.
\begin{table}
\begin{tabular}{l|l|c|l|l}
\hline 
Channel ($a$) & Distribution & \# bins   &\hspace*{0.2cm} Data set & \hspace*{0.2cm}Int Lum  \\ [0mm]
\hline
$WW\rightarrow \ell^+\ell^{\prime -}+\sla{E}_T\; (0j)$
& $p^{\rm leading, lepton}_{T}$
& 3 & ATLAS 8 TeV, &20.3 fb$^{-1}$~\cite{Aad:2016wpd}\\[0mm]
$WW\rightarrow \ell^+\ell^{(\prime) -}+\sla{E}_T\; (0j)$
& $m_{\ell\ell^{(\prime)}}$ & 8 & CMS 8 TeV, &19.4 fb$^{-1}$~\cite{Khachatryan:2015sga}\\[0mm]
$WZ\rightarrow \ell^+\ell^{-}\ell^{(\prime)\pm}$ & $m_{T}^{WZ}$ & 6 &
  ATLAS 8 TeV,& 20.3 fb$^{-1}$~\cite{Aad:2016ett}\\[0mm]
$WZ\rightarrow \ell^+\ell^{-}\ell^{(\prime)\pm}+\sla{E}_T$ & $Z$
    candidate $p_{T}^{\ell\ell}$ & 10 & CMS 8 TeV, &19.6 fb$^{-1}$~\cite{Khachatryan:2016poo}\\[0mm]
$WW/WZ \to \ell \nu J$ & $p_T^J$ & 6 & ATLAS 8 TeV, & 20.2 fb$^{-1}$~\cite{ATLAS:2017pbb}\\[0mm]  
  \hline
$WZ \to \ell^+ \ell^- \ell^{\prime\pm}$ & {$M(WZ)$} & 7& CMS 13
       TeV,  &  137.2 fb$^{-1}$~\cite{CMS:2021lix} \\[0mm]
$WW \to \ell^+\ell^{(\prime)-}+ 0/1 j$   &$M(\ell^+\ell^{(\prime)-})$
                             &11 & CMS 13 TeV, & 35.9 fb$^{-1}$~\cite{CMS:2020mxy} \\[0mm]
  $W\gamma \to \ell \nu \gamma$ & $\frac{d^2\sigma}{dp_Td\phi}$ & 12
       & CMS 13 TeV, & 137.1 fb$^{-1}$~\cite{CMS:2021rym}
       \\[0mm]
  $WW\rightarrow e^\pm \mu^\mp+\sla{E}_T\; (0j)$
&  $m_T$ & 17 (15) & 
ATLAS 13 TeV, &36.1 fb$^{-1}$~\cite{Aaboud:2017gsl} \\[0mm]
$WZ\rightarrow \ell^+\ell^{-}\ell^{(\prime)\pm}$ 
&  $m_{T}^{WZ}$ & 6 
& ATLAS 13 TeV, &36.1 fb$^{-1}$~\cite{ATLAS:2018ogj} \\[0mm]
$ Z jj \to \ell^+\ell^- jj$ & $\frac{d\sigma}{d\phi}$ & 12 & ATLAS 13 TeV,
& 139 fb$^{-1}$~\cite{ATLAS:2020nzk} \\[0mm]
\hline
\end{tabular}
\caption{Diboson data from LHC used to constrain the dimension-six operators.
  For the $W^+W^-$ results from ATLAS Run 2 ~\cite{Aaboud:2017gsl} we
  combined the data from the last three bins into one to
  ensure gaussianity.}
\label{tab:diboson}
\end{table}
Generically, for most data samples, from the above publications, we
extract the observed event rates in each bin ($N^{a}_{i,\rm d}$), as
well as the background expectations ($N^{a}_{i,\rm bck}$) and the SM
predictions ($N^{a}_{i,\rm sm}$) for each channel.  An exception are
the results for the $W^\pm\gamma$ of CMS \cite{CMS:2021rym}, and $Zjj$
in ATLAS ~\cite{ATLAS:2020nzk} for which the experiments provide the
results in the form of reconstructed differential cross sections.
\smallskip

In our analyses, we simulate the results in the $W^+W^-$, $W^\pm Z$,
$W^\pm \gamma$, and $Zjj$ channels that receive contributions from TGC
and anomalous fermion pair couplings to gauge bosons using
\textsc{MadGraph5\_aMC@NLO}~\cite{Frederix:2018nkq} with the UFO files
for our effective lagrangian generated with
\textsc{FeynRules}~\cite{Christensen:2008py, Alloul:2013bka}.  We
employ \textsc{PYTHIA8}~\cite{Sjostrand:2007gs} to perform the parton
shower and hadronization, while the fast detector simulation is
carried out with \textsc{Delphes}~\cite{deFavereau:2013fsa}.  Jet
analyses were performed using
\textsc{FASTJET}~\cite{Cacciari:2011ma}.\smallskip

In order to account for higher order corrections and additional
detector effects we obtain the SM diboson cross section in the
fiducial region defined by the experimental collaborations, requiring
the same cuts and isolation criteria employed in the experimental
studies, and normalize our results bin by bin to the experimental
collaboration predictions for the kinematic distributions under
consideration.  Then we apply these correction factors to our
simulated contributions to diboson distributions from the relevant
dimension--six operators.  \smallskip

These predictions are statistically confronted with the LHC Runs 1 and
2 data by constructing a binned log-likelihood function based on the
data contents of the different bins in the kinematic distribution of
each channel. In addition to the statistical errors, we incorporate
the systematic and theoretical uncertainties adding them in quadrature
and assuming some partial correlation among them which we estimate
with the information provided by the experiments.  In particular, some
of the experimental publications contain analysis of their EWDB data
in the framework of some anomalous TGC's. In those cases we can
compare the allowed region of parameter space obtained with our
constructed likelihood function for that data sample with that of the
experiment and we can use that comparison to fine-tune our treatment
of the systematic uncertainties and correlations to match the results
of the experimental analysis as closely as possible.  Altogether we
build the function $\chi^2_{\rm EWDBD}$ summing the $\chi^2$
constructed for the analysis of each sample in
Table~\ref{tab:diboson}. Then combining it with the EWPD bounds we
define
\begin{equation}
\chi^2_{\rm EWPD+EWDBD}(\fb, \fw, \fwww, \fbw, \fpone, \fqthree, \fqone,
\fur, \fdr, \fud, \fer,\fllll) \;\;.
\label{eq:chi2ewtgc}
\end{equation}
\begin{table} 
\begin{tabular}{lccccc}
\hline 
Source    & DATA FORMAT & ANALYSIS  & Int.Luminosity (fb$^{-1}$) &\hspace*{0.3cm} \# Data points
\\
\hline
ATLAS+CMS at 7 \& 8 TeV~\cite{Khachatryan:2016vau} [Table 8, Fig 27]
& SS  & SS \& STXS &  5 \& 20 & 20+1
\\
%ATLAS at 13 TeV~\cite{ATLAS:2018doi} [Figs. 6,7] & SS& 79.8 & 9  
%\\
ATLAS at 8 TeV~\cite{Aad:2015gba} ($\gamma Z$) &SS & SS \& STXS & 20 & 1 
\\\hline
%  ATLAS at 13 TeV~\cite{Aaboud:2017uhw} ($\gamma Z$) &SS & 36.1 & 1 
%  \\
%ATLAS at 13 TeV~\cite{ATLAS-CONF-2018-026} ($\mu^+\mu^-$) &SS & 36.1 & 1 
%\\
ATLAS at 13 TeV~\cite{ATLAS:2020qdt} [Figs. 7,20] & SS & SS & 36.1--139 & 9
\\
ATLAS at 13 TeV~\cite{ATLAS:2020qcv} ($\gamma Z$) &SS  & SS \& STXS & 139 & 1
  \\
ATLAS at 13 TeV~\cite{ATLAS:2020fzp} ($\mu^+\mu^-$) & SS & SS \& STXS & 139 & 1\\
ATLAS at 13 TeV ~\cite{ATLAS:2020naq}($\gamma\gamma, 4\ell,b\bar{b}$)
& STXS &STXS & 139 & 43
  \\
  ATLAS at 13 TeV~\cite{ATLAS:2020qdt} [Figs. 5,6]  & SS & STXS  & 36.1--139 & 7
\\
CMS at 13 TeV~\cite{CMS:2020gsy} [Table 5] &SS &  SS & 35.9--137 & 23 
\\  
CMS at 13 TeV~\cite{CMS:2020omd}
($\gamma\gamma$) & STXS & STXS &137 & 24
  \\
  CMS at 13 TeV~\cite{CMS:2021ugl}
  ($4 \ell$) & STXS & STXS&137 & 19
  \\
  CMS at 13 TeV~\cite{CMS:2020dvp} ($\tau^+\tau^-$) & STXS & STXS& 137&11
  \\
  CMS at 13 TeV~\cite{CMS:2021ixs} ($W^+W^-$) & STXS & STXS& 137&4
  \\
  CMS at 13 TeV~\cite{CMS:2020gsy} [Table 5] &SS &  STXS & 35.9--137 & 12 \\
  \hline
\end{tabular}
\caption{Higgs data used to compute $\chi^2_{\rm Higgs}$.  In the
  column labeled ``DATA FORMAT SS'' stands for total signal strength
  and STXS for simplified template cross section. The column marked
  ``ANALYSIS'' specifies in which of the two analyses we perform the
  data is included. In that column, ``SS'' labels the analysis which
  includes only total signal strengths for all channels and ``STXS''
  labels the analysis which includes instead the kinematic information
  in the form of simplified template cross sections for those channels
  for which it is available.}
  \label{tab:HiggsData}
\end{table}
In what respects Higgs results in our studies, we use the available
Higgs data depicted in Table~\ref{tab:HiggsData}.  Notice that
Ref.~\cite{ATLAS:2020naq} summarizes the ATLAS results on
$H\to\gamma\gamma$~\cite{ATLAS:2020pvn},
$H\to4\ell$~\cite{ATLAS:2020rej} and
$H\to b\bar{b}$~\cite{ATLAS:2020fcp} and the detailed statistics
information can be extracted from these references. \smallskip

In order to assess the effects of including the kinematic information
we have performed two different analyses of the Higgs results.  In a
first one we only include the data on the total signal strengths.  We
label that analysis SS. It includes the results labeled as ANALYSIS SS
in Table~\ref{tab:HiggsData} which corresponds to 22 data points for
Run 1 and 11(ATLAS)+23(CMS)=34 data points for Run 2 for a total of 56
data points. \smallskip

On the second analysis, labeled as ANALYSIS STXS, we make use of all
the available information on the kinematic distributions as provided
in the form of simplified template cross sections. For those channels
and luminosities for which the STXS data is not available we use the
corresponding data as total signal strength.  In this form the STXS
analysis includes the same 22 data point for Run 1 and 45 STXS (58)
points for the ATLAS (CMS) Run 2, supplemented with the 7 (12) SS for
the channels for which no STXS data is available. So in total de STXS
analysis includes 144 data points.  It is important to notice that the
STXS results of ATLAS for the different channels are all presented in
Ref.~\cite{ATLAS:2020naq} where a full correlation matrix between the
different STXS bins for the different channels is provided. On the
contrary the different STXS results from CMS are given in different
publications Refs.~\cite{CMS:2020omd, CMS:2021ugl, CMS:2020dvp,
  CMS:2021ixs} for the different channels and no correlations between
bins for different channels is available. So in our analysis we have
set those to to zero.  \smallskip

We obtained the theoretical predictions for the Higgs production by
gluon fusion in the channels tagged as STXS in
Table~\ref{tab:HiggsData} using
\textsc{MadGraph5\_aMC@NLO}~\cite{Hirschi:2015iia} with the SMEFT@NLO
UFO files~\cite{Degrande:2020evl}.  Notice that we took into
  account not only the contribution to this process from $\ogg$ but
  also the ones coming from $\optwo$, $\ot$ and $\otg$.   Moreover,
the STXS 1.2 classification was performed using
\textsc{Rivet}~\cite{Buckley:2010ar}.  \smallskip

The statistical comparison of our effective theory predictions with
the LHC results including those two data samples is made by means of a
$\chi^2$ analysis performed in the 21 dimensional statistical function
and adding the information from EWPD and EWDBD.  Furthermore, the
operator $\otg$ contributes not only to the gluon-fusion Higgs
production~\cite{Deutschmann:2017qum, Grazzini:2018eyk}, but also to
the top production~\cite{Buckley:2015lku, Aguilar-Saavedra:2018ksv,
  Brivio:2019ius, CMS:2019zct, Bissmann:2020mfi, Ethier:2021bye}which
provides independent information.  We include such information on this
operator in the form of a gaussian bias built with the result of the
full global fit to top-quark physics performed in
Ref.~\cite{Brivio:2019ius}.
\begin{equation}
  \chi^2_{\rm bias,top} (\ftG)
  =\frac{\displaystyle\left(
      \frac{\displaystyle\ftG}{\displaystyle \Lambda^2}-0.5\right)^2}{(0.22)^2}\;.
\label{eq:bias}
\end{equation}

Altogether we define the $\chi^2$ for our global analysis as:
\begin{eqnarray}
  \hskip -1cm
  \chi^2_{\rm Global\; SS (STXS)}=&&
  \chi^2_{\rm EWPD+EWDBD}
  (\fb, \fw, \fwww, \fbw, \fpone, \fqthree, \fqone,\fur, \fdr, \fud, \fer,\fllll) \\\nonumber
 && + \chi^2_{\rm HIGGS\; SS (STXS)}
  (\fb, \fw, \fbb,\fww,\fbw,\fgg, \fpone 
  \fptwo,\fqthree, \fqone, \fur, \fdr, \fud, \\\nonumber
  && \hspace*{10cm} \fer,\ftG,\fbo,\ft,\fta,\fm)\\\nonumber
  &&+\chi^2_{\rm bias,top} (\ftG) \;\;.
\label{eq:chiglobal}
\end{eqnarray}

%%%%%%%%%%%%%%%%%%%%%%%%%%%%%%%%%%%%%%%%%%%%%%%%%%%%%%%%%%%%%%%%%%%%%%
\section{Effective Field Theory Results}
\label{sec:results}

Figures~\ref{fig:fer_r2}--\ref{fig:corr-phi2bt} and \ref{fig:globstxs}
depict $\Delta\chi^2$ profiles for the dimension-six Wilson
coefficients, where we marginalized with respect to all undisplayed
parameters.  We show the results for four different combinations of
data sets:
\begin{itemize}
\item EWPD: $\Delta\chi^2_{\rm EWPD}$ which constrains the 8
  coefficients in $\Delta{\cal L}_{\rm eff}^{\rm EWPD}$,
  Eq.~\eqref{eq:leff-ewpd}.  They are given by the green lines in
  Fig.~\ref{fig:fer_r2}, which have been obtained considering only the
  contributions to the observables that are ${\cal O}(\Lambda^{-2})$;
  see Ref.~\cite{Corbett:2017qgl}. We also assess the impact of
    relaxing the flavor universality hypothesis by considering a
    different coupling to the third family whose result is displayed
    in Fig.~\ref{fig:fer_r2_nofam}. 
\item EWPD+EWDBD: $\Delta\chi^2_{\rm EWPD+EWDBD}$ which limits the 12
  coefficients in
  $ \Delta {\cal L}_{\rm eff}^{\rm EWPD}+\Delta {\cal L}_{\rm
    eff}^{\rm TGC}$, Eqs.~\eqref{eq:leff-ewpd}
  and~\eqref{eq:leff-tgc}. In the EWDBD analysis we considered
  contributions to the observables up to ${\cal O}(\Lambda^{-2})$ and
  up to ${\cal O}(\Lambda^{-4})$.  Our results are displayed in
  Fig.~\ref{fig:tgv_r2}.
\item GLOBAL: $\Delta\chi^2_{\rm Global\;ANALISIS}$
  \eqref{eq:chiglobal} which constrains the 21 coefficients in
  ${\cal L}_{\rm eff}$ in
  Eqs.~\eqref{eq:leff-ewpd}--\eqref{eq:leff-h}. As described in
  previous section we consider two different sets of HIGGS data, one
  which includes only the total signal strengths, ANALISIS$\equiv$SS,
  and one which includes also the information on the kinematic
  distributions, ANALYSYS$\equiv$STXS (see Table
  ~\ref{tab:HiggsData}). For each case we perform two analyses: one in
  which the theoretical predictions are expanded to
  ${\cal O}(\Lambda^{-2})$, and another one including the predictions
  up ${\cal O}(\Lambda^{-4})$.  Different projections of the
  corresponding $\Delta\chi^2_{\rm Global\;ANALISIS}$ are shown in
  Figs.~\ref{fig:fer_r2}--\ref{fig:corr-phi2bt}, and
  Fig.~\ref{fig:globstxs} summarizes the results for the STXS
  analysis.
\end{itemize}

%%%%%%%%%%%%%%%%%%%%%%%%%%%%%%%%%%%%%%%%%%%%%
\subsection{Lessons from the EWPD}
\label{sec:ferm}

\begin{figure}[h!]
  \centering
  \includegraphics[width=0.85\textwidth]{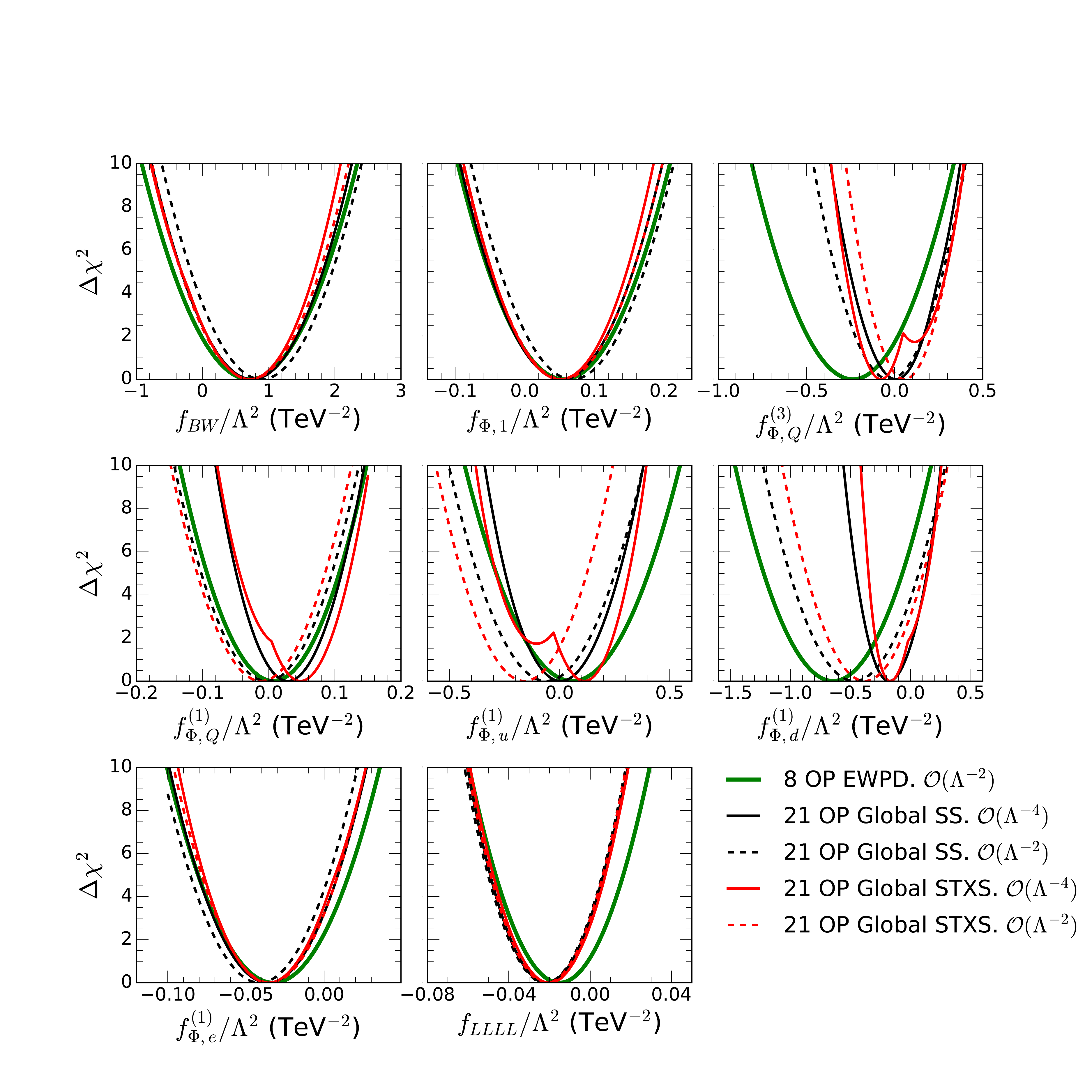}
  \caption{$\Delta \chi^2$ as a function of the Wilson coefficients
    $\fbw/\Lambda^2$, $\fpone/\Lambda^2$, $\fqone/\Lambda^2$,
    $\fqthree/\Lambda^2$, $\fur/\Lambda^2$, $\fdr/\Lambda^2$,
    $\fer/\Lambda^2$, and $\fllll/\Lambda^2$, as indicated in the
    panels after marginalizing over the remaining fit parameters. The
    green solid line stands for the fit of the EWPD that constrains
    only eight of twenty one Wilson coefficients in
    Eq.~(\ref{eq:leff}). The red solid (dashed) line represents the
    twenty-one-parameter fit to the LHC Run 1 and 2 data including the
    STXS Higgs data and working at order $1/\Lambda^4$ ($1/\Lambda^2$)
    approximation. For comparison, we present the corresponding
    results for the global fit using the SS Higgs data (black solid
    and dashed lines).}
  \label{fig:fer_r2}
\end{figure}

We start our analysis focusing on the dimension--six operators in
Eq.~\eqref{eq:leff-ewpd} that impact directly EWPD. Our results are
presented in Fig.~\ref{fig:fer_r2}.  The bounds imposed by the EWPD
analysis are represented by the green lines while the limits coming
from the global SS (STXS) analyses are given by the black (red) lines
including up to ${\cal O}(\Lambda^{-4})$ and up to
${\cal O}(\Lambda^{-2})$ terms as full and dashed lines respectively.
As it is well-known, EWPD leads to very stringent constraints on
fermion-gauge interactions and on new {\sl oblique} corrections to the
gauge boson propagators.  However, as illustrated in the figure, with
the accumulated LHC statistics, EWDBD and Higgs results already
contribute to tighten the EWPD bounds for some of the coefficients
despite the larger 21 parameter space.  \smallskip
%%%%%%%%%%%%%%%%%%%%%%%%%%%%%%%%%%%%%%%%%%%%%%%%%%%%%%%%%%%%%%%%
%%%%%%%%%%%%%%%%%%%%%%%%%%%%%%%%%%%%%%%%%%%%%%%%%%%%%%%%%%%%%%%%

In the upper left and central panels of Fig.~\ref{fig:fer_r2}, we find
the $\Delta\chi^2$ dependence on $\fbw/\Lambda^2$ and
$\fpone/\Lambda^2$ which correspond to the $S$ and $T$ oblique
corrections respectively.  For these coefficients, the EWPD limits are
only slightly improved just by the global analysis including the STXS
Higgs data when the theoretical predictions include the quadratic
contributions of the Wilson coefficients.  \smallskip

\begin{figure}[h!]
  \centering
  \includegraphics[width=0.65\textwidth]{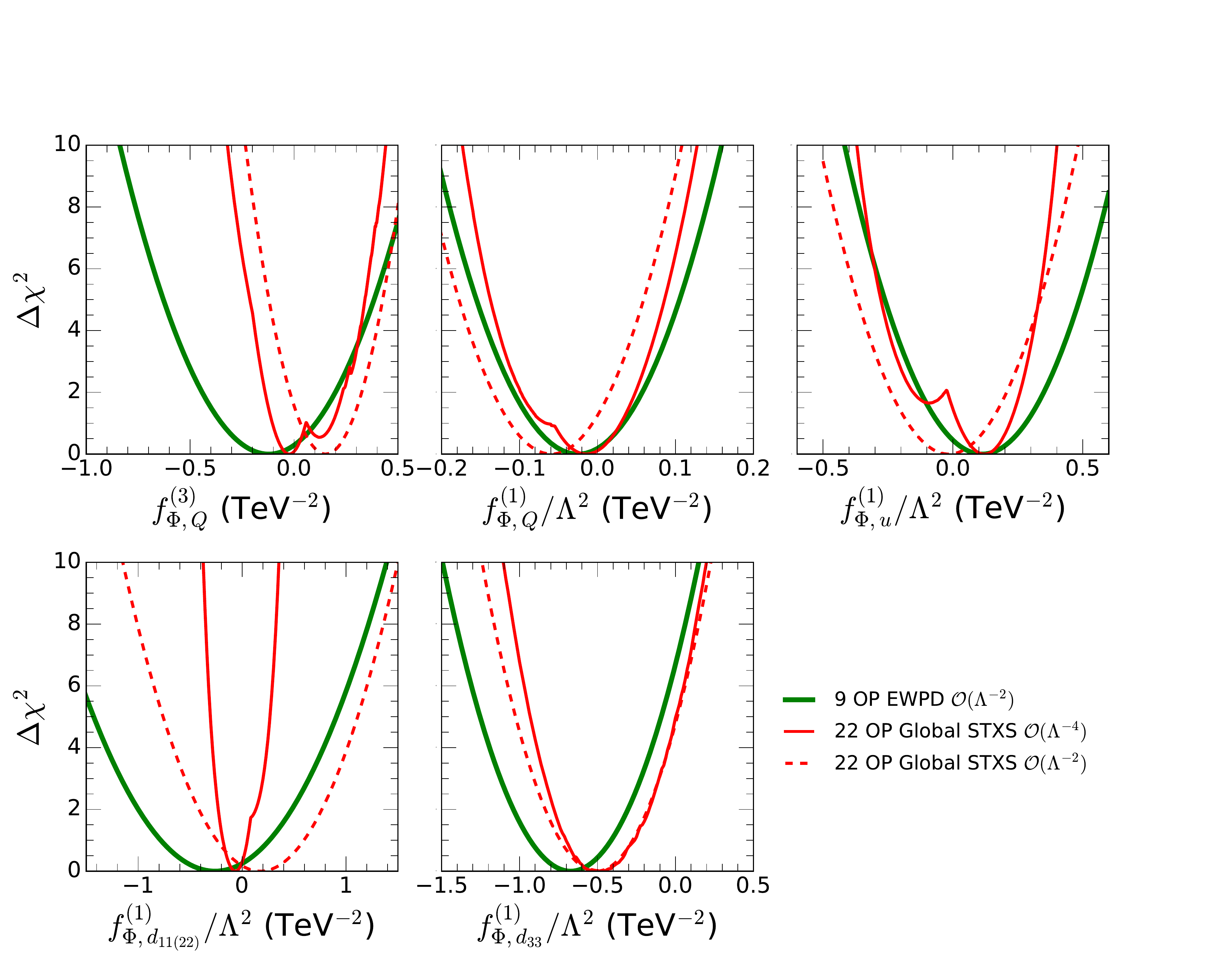}
  \caption{$\Delta \chi^2$ as a function of the Wilson coefficients
    contributing to the gauge couplings of the quarks for
    non-universal right-handed quark coupling of the bottom quark,
    this is for
    $f^{(1)}_{\Phi,d_{1}}=f^{(1)}_{\Phi,d_{2}}\neq
    f^{(1)}_{\Phi,d_{3}}$.  green solid line stands for the fit of the
    EWPD that constrains only nine of twenty two Wilson coefficients.
    The red solid (dashed) line represents the twenty-one-parameter
    fit to the LHC Run 1 and 2 data including the STXS Higgs data and
    working at order $1/\Lambda^4$ ($1/\Lambda^2$) approximation.}
  \label{fig:fer_r2_nofam}
\end{figure}

Turning to the quark gauge couplings, the right upper (left central)
panel in Fig.~\ref{fig:fer_r2} present the $\Delta\chi^2$ dependence
on the coefficient $\fqthree/\Lambda^2$ ($\fqone/\Lambda^2$), which
modifies the couplings of left-handed quarks to $Z$ and $W$ (only $Z$)
bosons.  The middle and right central panels correspond to the
dependence on $\fur/\Lambda^2$ and $\fdr/\Lambda^2$ which give
corrections to the $u_R$ and $d_R$ couplings to $Z$ respectively.  The
inclusion of the LHC observables leads to a small but not negligible
improvement on the determination of $\fur/\Lambda^2$ at linear order,
while it only affects the EWPD bounds on $\fqone/\Lambda^2$ when
${\cal O}(\Lambda^{-4})$ terms are included in the LHC analysis.  The
largest effect of the combination with the LHC observables occurs for
$f^{(3)}_{\Phi d}/\Lambda^2$ and $\fqthree/\Lambda^2$.  In particular
the EWPD analysis favors non-vanishing values for
$f^{(1)}_{\Phi d}/\Lambda^2$ at 2$\sigma$, a result driven by the
2.7$\sigma$ discrepancy between the observed $A_{\rm FB}^{0,b}$ and
the SM. On the contrary, the inclusion of the LHC data gives rise to a
shift towards zero of $f^{(1)}_{\Phi d}/\Lambda^2$, reducing the
tension with the SM.  This behavior was already observed in
Refs.~\cite{Alves:2018nof, daSilvaAlmeida:2018iqo} but we find now
that with the accumulated statistics, LHC also is able to provide
relevant constraints when including the effect of dimension-six
operators only at ${\cal O}(\Lambda^{-2})$. As the LHC observables
included are mostly sensitive to gauge couplings of the light quarks
in the parton distribution functions, one expects that this result
relies upon the assumption that the operators involving fermion
couplings to gauge bosons are generation independent.  To test this,
we perform an analysis in which we drop this assumption for the
operator $f^{(1)}_{\Phi d}/\Lambda^2$ and allow
$f^{(1)}_{\Phi,d_{11}}=f^{(1)}_{\Phi,d_{22}}\neq
f^{(1)}_{\Phi,d_{33}}$ \footnote{If one relaxes the assumption of
  generation independent quark couplings the EWPD cannot constraint
  all quark couplings because there is not enough information in the
  observables considered to resolve the contributions of the two first
  generations.  Furthermore, for the third generation of quarks only
  $f^{(1)}_{\Phi,d_{33}}$ and the linear combination
  $4\,f^{(1)}_{\Phi,Q_{33}} + f^{(3)}_{\Phi,Q_{33}}$ contributes
  independently to the Z and W observables.}. The results are shown in
Fig.~\ref{fig:fer_r2_nofam} where we display the bounds imposed by the
EWPD on the relevant quark operators with those from the global STXS
analysis including up to ${\cal O}(\Lambda^{-2})$ and up to
${\cal O}(\Lambda^{-4})$ terms. Comparing with the corresponding
panels in Fig.~\ref{fig:fer_r2} we see that the combination of EWPD
with the LHC observables results in the quoted improvement on the
bounds for those operators contributing to the light quark
couplings. Conversely $f^{(1)}_{\Phi,d_{33}}$ is only marginally
affected by the inclusion of the EWDBD and Higgs data and its best fit
remains no-zero at $\sim 2\,\sigma$ in the global analysis. \smallskip

%%%%%%%%%%%%%%%%%%%%%%%%%%%%%%%%%%%%%%%%%%%%%%%%%%%%%%% 
%%%%%%%%%%%%%%%%%%%%%%%%%%%%%%%%%%%%%%%%%%%%%%%%%%%%%%%

Finally, the lower panels of Fig.~\ref{fig:fer_r2} display the
marginalized $\Delta\chi^2$ for $\fer/\Lambda^2$ and
$\fllll/\Lambda^2$. The global analyses do not lead to significant
improvement on the determination of these couplings. This is expected
since the operator $\oer$ modifies the $Z$ coupling to right-handed
leptons which were very precisely tested at LEP. On the contrary, it
enters the LHC observables only via its contribution to the decay rate
of the $Z$ boson to leptons in some of the final states considered.
In other words, the dominant dependence of the global analysis on
these coefficients still resides in the EWPD. The main effect of the
inclusion of the LHC results is indirect via the restriction of the
allowed range of variation of the other coefficients entering in the
EWPD analysis.  \smallskip

%%%%%%%%%%%%%%%%%%%%%%%%%%%%%%%%%%%%%%%%%%%%%
\subsection{Triple anomalous gauge couplings constraints}
\label{sec:tgc}

\begin{figure}[h!]
  \centering
  \includegraphics[width=0.8\textwidth]{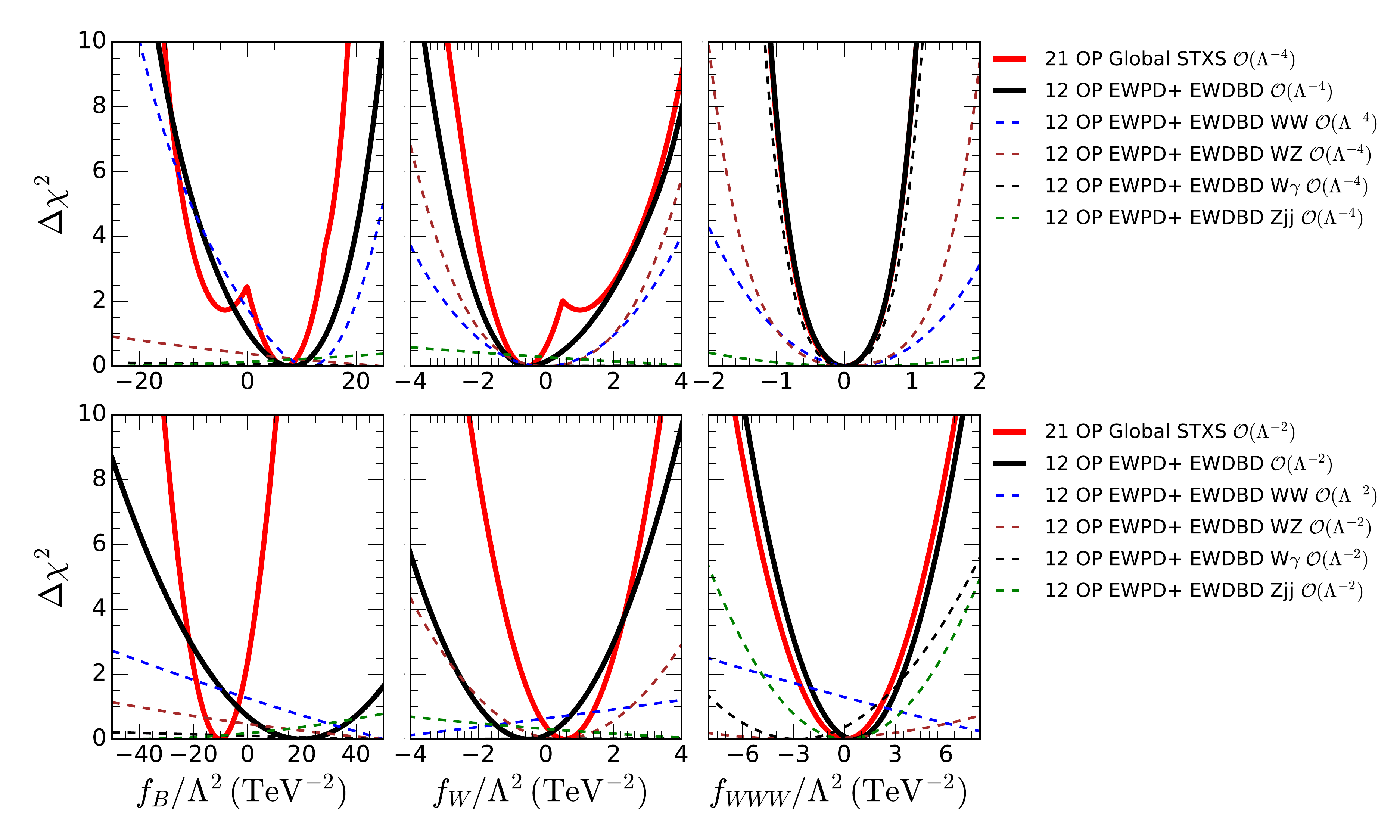}
  \caption{$\Delta \chi^2$ dependence on the $f_B/\Lambda^2$ (left
    panels), $f_{W}/\Lambda^2$ (central panels) and
    $f_{WWW}/\Lambda^2$ (right panels) parameters after the
    marginalization over the 11 (20) undisplayed fit parameters for the
    analysis of EWDBD+EWPD (global STXS) data as labeled in the figure.
    The upper panels show the results
    of our analysis using up to ${\cal O}(\Lambda^{-4})$ terms in the Wilson
    coefficients while the lower ones retained only the
    ${\cal O}(\Lambda^{-2})$
    terms. }
  \label{fig:tgv_r2}
\end{figure}

We present our results on the Wilson coefficients of the TGC operators
$\ob$, $\ow$ and $\owww$ in Fig.~\ref{fig:tgv_r2}. We first consider
the twelve-dimensional parameter space
\[
\left \{
\fb, \fw, \fwww, \fbw, \fpone, \fqthree, \fqone,
\fur, \fdr, \fud, \fer,\fllll
\right \}
\]
and include the EWDBD and EWPD.  The upper (lower) panels contain the
one--dimensional $\Delta\chi^2$ distributions after marginalization
over the 11 undisplayed coefficients and they were obtained using the
up to quadratic (linear) contributions of the 12 Wilson
coefficients. Each panel shows as dashed lines the impact of the
different diboson data sets and as full black line the combined
EWPD+EWDBD analysis.  For comparison we also show the corresponding
projections obtained from the global STXS fit performed in the the 21
parameter space (red lines). \smallskip

In what respects $\fb/\Lambda^2$ we find that EWPD+EWDBD provides only
lose constraints on this parameter at ${\cal O}(\Lambda^{-2})$ (see
black line in the lower left panel of Fig.~\ref{fig:tgv_r2}).
Comparing with the black line in the upper left panel of the figure we
see how the inclusion of the ${\cal O}(\Lambda^{-4})$ terms leads to
better bounds, with the $WW$ channel (blue dashed line) playing the
dominant role.  This is so even though the integrated luminosity is
larger for the $W\gamma$ data than for $WW$, because the $W\gamma$
data depends on the combination $(\fb/\Lambda^2+\fw/\Lambda^2)$ which
allows for cancellations between $\fb/\Lambda^2$ and
$\fw/\Lambda^2$. Similarly the $WZ$ channel also depends on
$\fb/\Lambda^2$ in another linear combination with $\fw/\Lambda^2$ and
furthermore its coefficient is suppressed by a factor $\tan^2\theta_W$
with respect to the $\fw/\Lambda^2$.  From these panels we also
conclude that $\fb/\Lambda^2$ is much more strongly constrained by the
global analysis (red lines) than by the EWPD+EWDBD.  This is so
because $\ob$ contributes to both Higgs production and decay.  This
fact was observed before in Ref.~\cite{Corbett:2013pja}.  But we find
now that with the large Run 2 luminosity and the STXS information, the
constraints on $\fb/\Lambda^2$ are substantially improved even when
including the dimension-six parameters only up to
${\cal O}(\Lambda^{-2})$.  \smallskip

The $\Delta\chi^2$ distributions for $\fw/\Lambda^2$ are displayed in
the middle panels where we see that the dominant diboson channels
constraining $\fw/\Lambda^2$ are the $WW$ and $WZ$. Here too,
$W\gamma$ production is not able to produce any sensible bound on
$\fw/\Lambda^2$ due to its (anti-)correlation with $\fb/\Lambda^2$
previously commented.  In this case the limits obtained at
${\cal O}(\Lambda^{-2})$ are already rather stringent and they are
significantly improved by the global analysis.\smallskip

\begin{figure}[h!]
  \centering
  \includegraphics[width=0.45\textwidth]{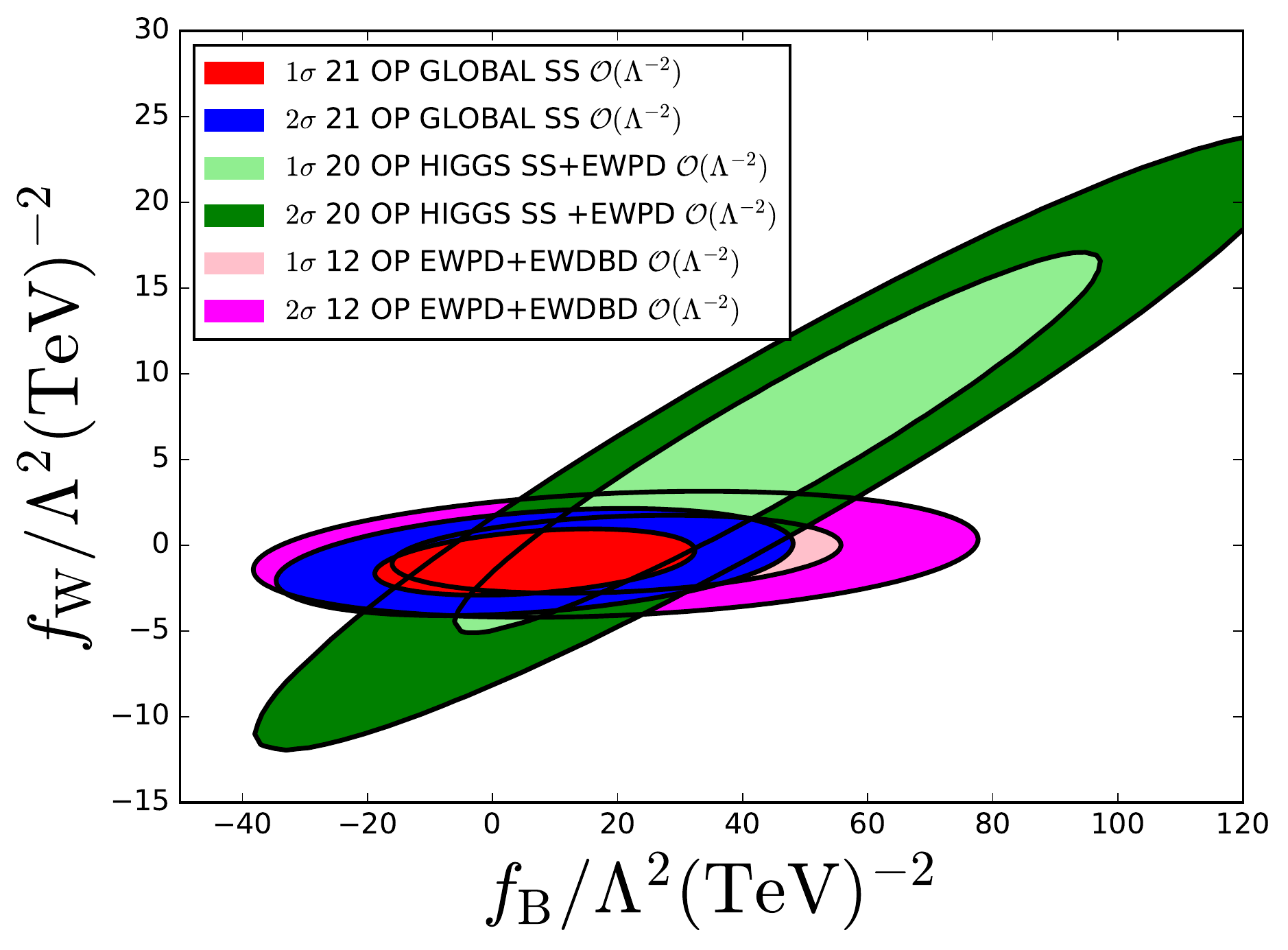}
  \includegraphics[width=0.45\textwidth]{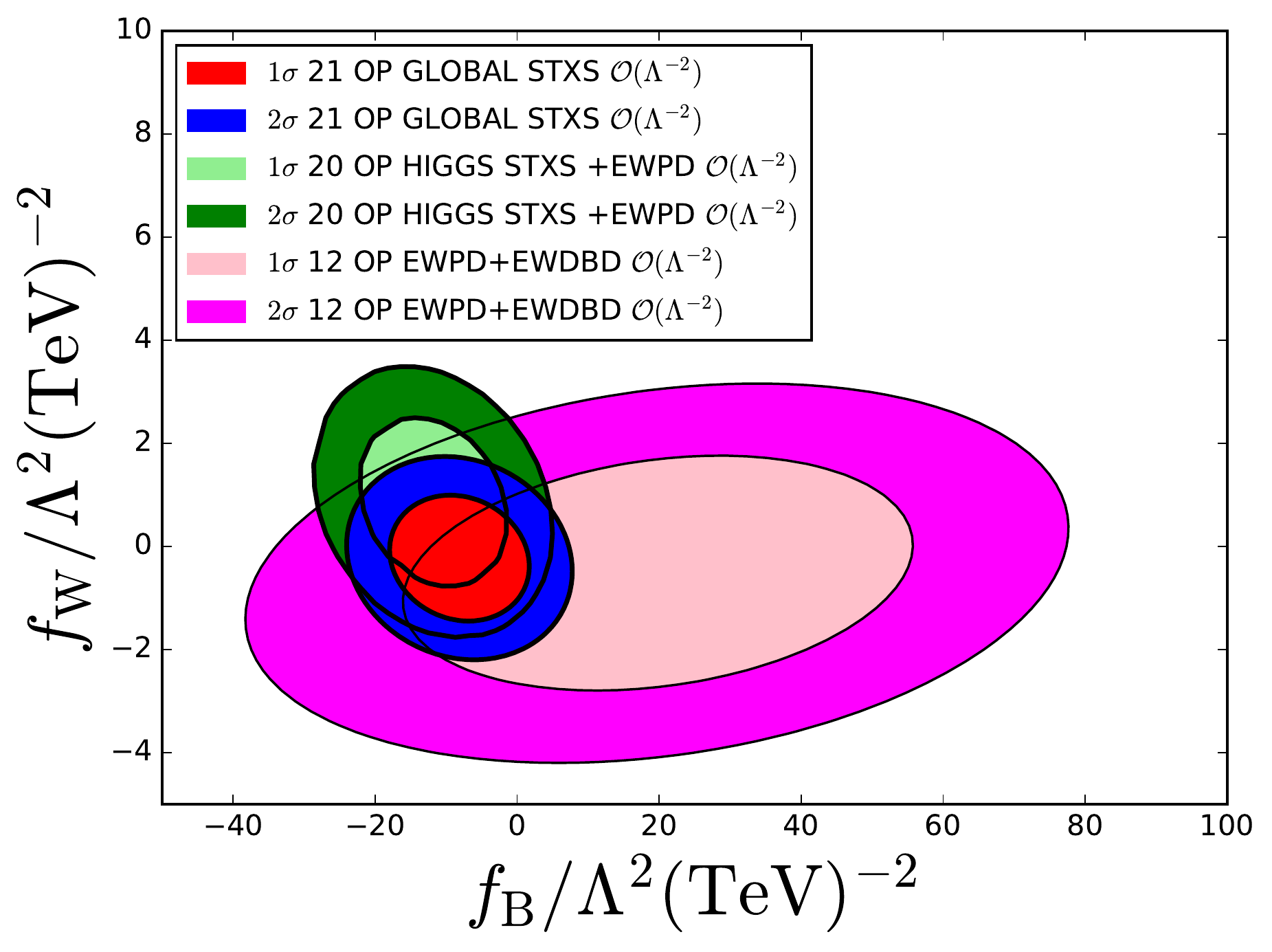}
  \caption{ $1\sigma$ and $2\sigma$ (2dof) allowed regions on the
    $\fb/\Lambda^2 \times \fw/\Lambda^2$ plane obtained from the
    analysis of different combinations of data samples as labeled in
    the figure.  All results use predictions up to the
    ${\cal O}(\Lambda^{-2})$ terms in the Wilson coefficients and have
    been marginalized over the undisplayed ones.}
  \label{fig:2dfbfw}
\end{figure}

In order to show the importance of the Higgs data in constraining
$\fb/\Lambda^2$ and $\fw/\Lambda^2$ we present in
Fig.~\ref{fig:2dfbfw} the $1\sigma$ and $2\sigma$ allowed regions of
the plane $ \fb/\Lambda^2 \times \fw/\Lambda^2$ obtained with
different data samples. These results were obtained using only the
contributions up to ${\cal O}(\Lambda^{-2})$ on the Wilson
coefficients and the SS (STXS) Higgs data on the left (right)
panel. As we can see the EWPD+EWDBD data sets lead to a very large
allowed region (pink regions). The constraints coming from the SS
Higgs data (shown in the green regions on the left panel in
combination with the EWPD ) are also loose but they are important to
improve the limits when combined also with the EWDBD.  But most
interestingly, as seen in the right panel, using \emph{only} the Higgs
STXS analysis in combination with EWPD (right panel), is able to place
much stronger bounds on $\fw/\Lambda^2$ and $\fb/\Lambda^2$ ; compare
the green and pink regions.  Clearly, the global fit on $\ob$ is
dominated by the data samples in the Higgs STXS analysis while the
constraints on $\ow$ receives comparable contributions from Higgs STXS
and EWDBD. \smallskip

At last, the right panels of Fig.~\ref{fig:tgv_r2} contain the
dependence of the marginalized $\Delta\chi^2$ on $\fwww/\Lambda^2$. In
contrast to $\fb/\Lambda^2$ and $\fw/\Lambda^2$, the channel $W\gamma$
plays a significant role in constraining $\fwww/\Lambda^2$ due to the
use of kinematic distributions specially chosen to avoid the
cancellation of the $1/\Lambda^2$ contribution
~\cite{Azatov:2017kzw,Azatov:2019xxn}.  In fact, we can see from the
lower right panel of this figure that the most important contributions
to the EWPD+EWDBD analysis originates from those channels ($W\gamma$
and $Zjj$).  Also, as expected, the global analysis has barely any
additional impact on limiting $\owww$ since this operator does not
contribute to the Higgs observables.  \smallskip

%%%%%%%%%%%%%%%%%%%%%%%%%%%%%%%%%%%%%%%%%%%%%%%%%%%%%%%%%%%%%%%%
%%%%%%%%%%%%%%%%%%%%%%%%%%%%%%%%%%%%%%%%%%%%%%%%%%%%%%%%%%%%%%%%

%%%%%%%%%%%%%%%%%%%%%%%%%%%%%%%%%%%%%%%%%%%%%%%%%%%%%%%%%%%%%%%%
%%%%%%%%%%%%%%%%%%%%%%%%%%%%%%%%%%%%%%%%%%%%%%%%%%%%%%%%%%%%%%%%

%%%%%%%%%%%%%%%%%%%%%%%%%%%%%%%%%%%%%%%%%%%%%%%%%%%%%%%%%%%%%%%%%
\subsection{Higgs couplings}
\label{sec:higgs}

In order to probe for deviations from the SM predictions to the Higgs
couplings we performed four global fits including the effects of the
21 operators in Eqs.~\eqref{eq:leff-ewpd}--\eqref{eq:leff-h} under
different assumptions.  As mentioned in Sec.~\ref{sec:frame} in order
to access the importance of the newly available kinematic
distributions we made two analysis: one in which that information is
not included (Global SS) and another one in which it is (Global
STXS). And, as for EWDBD, we make two variants of the analysis, one
employing the theoretical predictions up to ${\cal O}(\Lambda^{-2})$
terms in the Wilson coefficients, and one including up to
${\cal O}(\Lambda^{-4})$ terms. \smallskip

\begin{figure}[h!]
\centering
\includegraphics[width=0.85\textwidth]{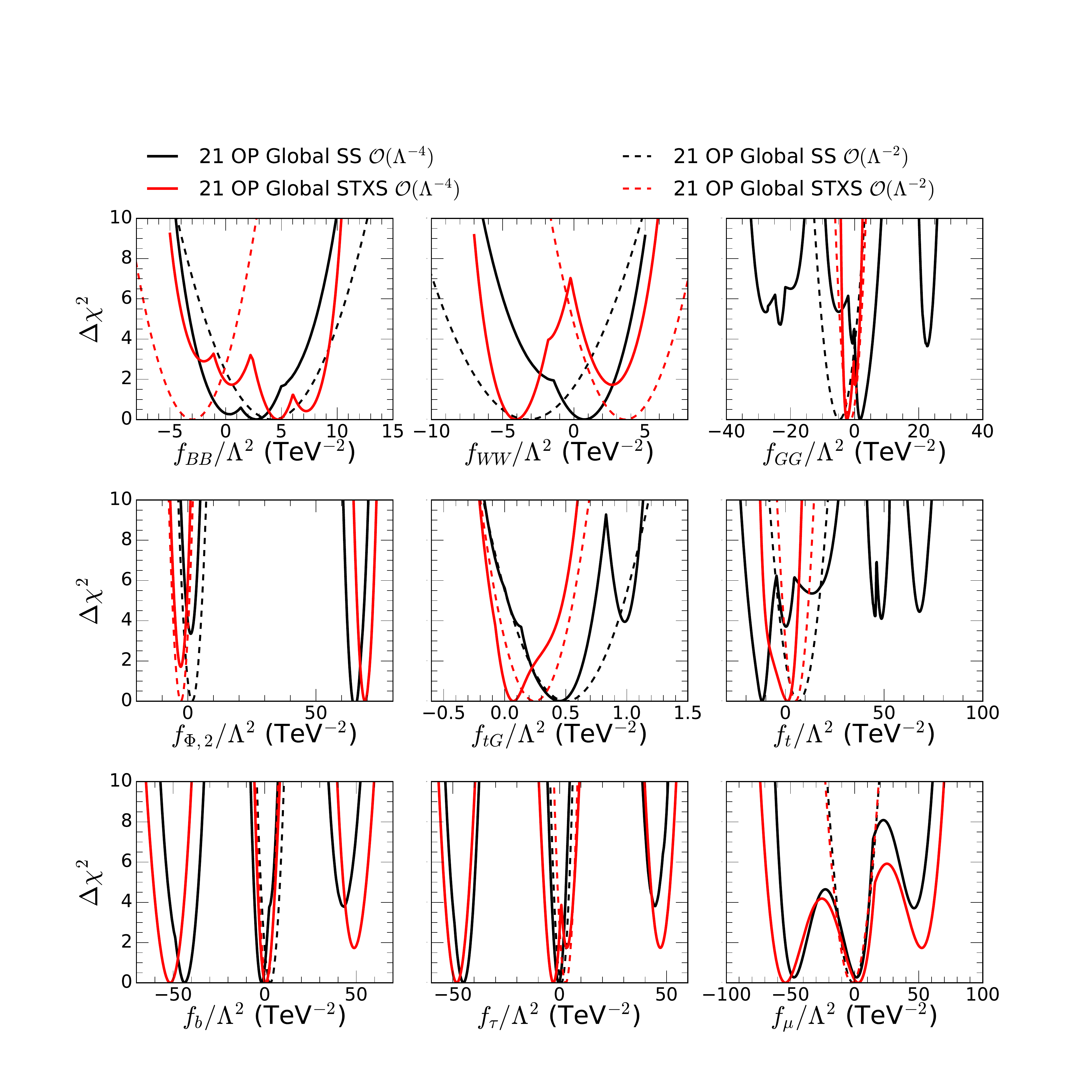}
\caption{Marginalized $\Delta\chi^2$ distributions as a function of the Wilson
  coefficients for the nine operators that only affect the Higgs physics 
  and for the four variants of the global analysis as labeled in the figure
  (see text for details).}
  \label{fig:bos_r2}
\end{figure}

We have discussed in the context of Figs.~\ref{fig:fer_r2}
and~\ref{fig:tgv_r2} the results from these global analysis for the 12
operators that contribute also to the EWPD and EWDBD, therefore, we
focus here on the nine operators not studied yet. We show in
Fig.~\ref{fig:bos_r2} the dependence of the marginalized
$\Delta\chi^2_{\rm global}$ on each of these nine Wilson coefficients
for the four analysis variants. \smallskip

%%%%%%%%%%%%%%%%%%%%%%%%%%%%%%%%%%%%%%%%%%%%%%%%%%%%%%%%%%%%%%%% 
%%%%%%%%%%%%%%%%%%%%%%%%%%%%%%%%%%%%%%%%%%%%%%%%%%%%%%%%%%%%%%%%

From Fig.~\ref{fig:bos_r2} we observe , as expected,
that for the analysis including
only the contributions up to ${\cal O}(\Lambda^{-2})$ (dashed lines)
there is a unique minimum in either the global SS or global STXS
analysis.  And comparing the red and black dashed lines we
  conclude that at ${\cal O}(\Lambda^{-2})$ the impact of the STXS
  observables in the overall picture amounts to an uncertainty
  reduction of 30--40\% for some of the bosonic operators as well as a
  shift in the allowed region in a few cases. \smallskip

Conversely, we also see in Fig.~\ref{fig:bos_r2} that for the analysis
including up to ${\cal O}(\Lambda^{-4})$ terms and the SS samples for
the Higgs data (solid black lines) all panels present some set of
(quasi)degenerated minima.  They are a direct reflect of the
(quasi)degeneracies in the Higgs couplings in
Eqs.~\eqref{eq:vert-hww}--\eqref{eq:vert-gluglu}. Comparing the solid
black and red lines we see the relevance of the kinematic
distributions in resolving some of these degeneracies. \smallskip

First, let us focus on the left central panel of Fig.~\ref{fig:bos_r2}
which depicts the $\Delta\chi^2$ distribution as a function of
$\fptwo/\Lambda^2$.  In the SS analysis at ${\cal O}(\Lambda^{-4})$,
there are two clearly almost degenerate minima associated with the
flip of sign of all Higgs couplings discussed below
Eq.~\eqref{eq:vert-hww} for $\fptwo/\Lambda^2 \simeq 65$
TeV$^{-2}$. In fact we find that the analysis show a slight preference
for this non-standard solution.  As seen in the figure, this is still
the case once the information of the STXS observables is included but
the $\Delta\chi^2$ of the SM-connected solution is reduced to
$\sim 1.3\;\sigma$. \smallskip

\begin{figure}[h!]
\centering
\includegraphics[width=0.75\textwidth]{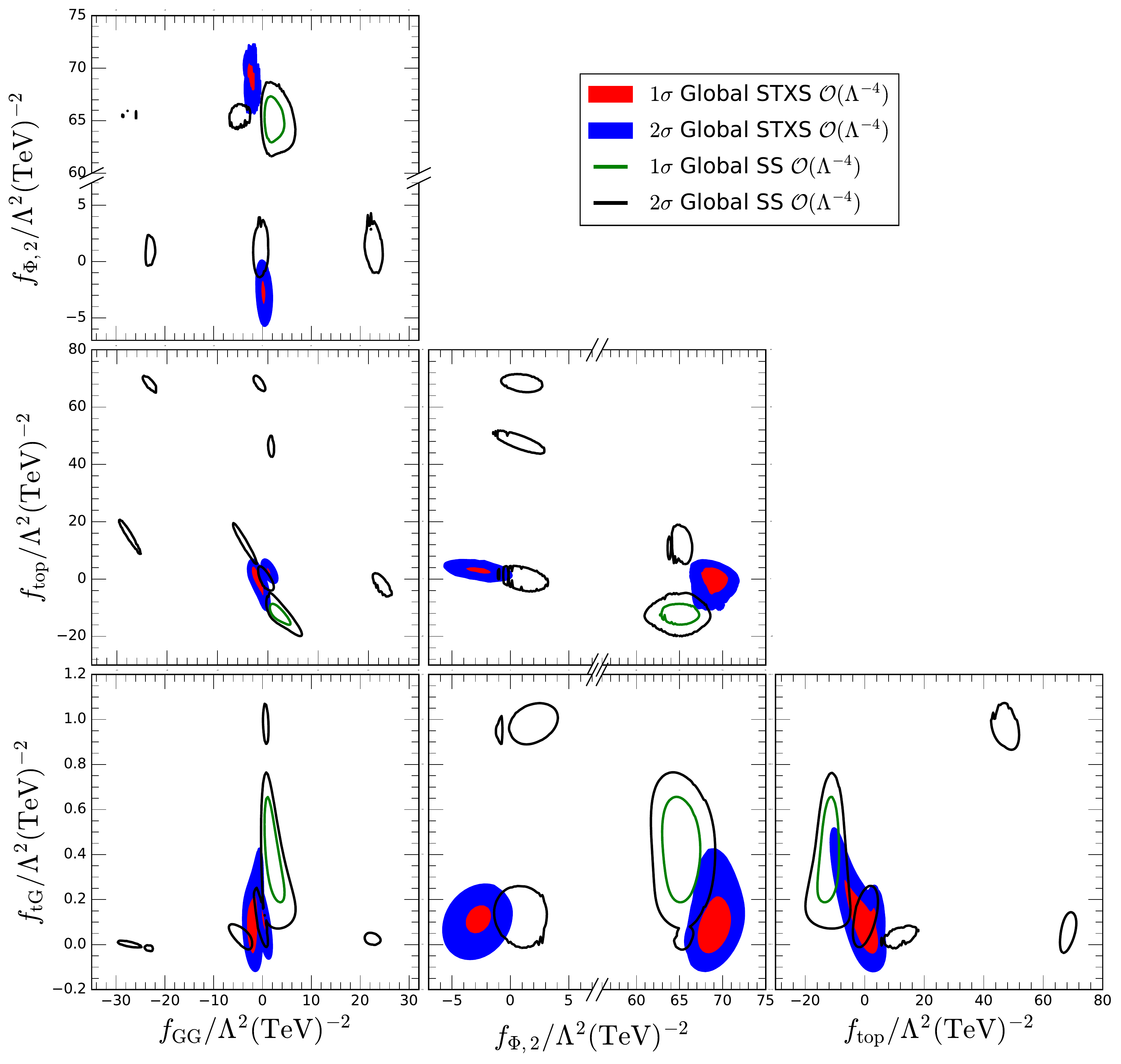}
\caption{$1\sigma$ and 95\% CL (2dof) allowed regions from the SS and
  STXS global analysis for the coefficients of the operators entering
  in the gluon-gluon Higgs vertex (Eq.~\eqref{eq:vert-gluglu}). 
  These results were
  obtained using terms up to  ${\cal O}(\Lambda^{-4})$
  of the theoretical prediction,
  and the color code indicates the used data set as labeled in the figure.}
  \label{fig:2dgluglu}
\end{figure}

The power of the kinematic distributions is particularly striking for
the coefficients $\fgg/\Lambda^2$ , $\ft/\Lambda^2$ and
$\ftG/\Lambda^2$ which, together with $\fptwo/\Lambda^2$, enter in the
effective gluon-gluon Higgs vertex and for which the marginalized
$\chi^2$ for the global SS analysis shown in the black curves in
Fig.~\ref{fig:bos_r2} present a complex structure of local minima.
This is further illustrated in Fig.~\ref{fig:2dgluglu}.  In this
figure we show the allowed regions from the global analysis performed
at ${\cal O}(\Lambda^{-4})$, projected over pairs of these parameters
(after marginalization over the 19 undisplayed parameters in each
panel).  We observe the complex structure of allowed regions in
$\fgg/\Lambda^2 \times \fptwo/\Lambda^2$,
$\fgg/\Lambda^2 \times \ft/\Lambda^2$ and
$\ftG/\Lambda^2 \times \fptwo/\Lambda^2$ that appear in the SS
${\cal O}(\Lambda^{-4})$ analysis~\footnote{We notice that the fact
  that the external bias on $\ftG$ from top observables
  (Eq.\eqref{eq:bias}) is not centered at zero further adds to the
  complex structure of local minima.}.  As inferred from the figure,
the inclusion of the Higgs kinematic distributions in the analysis is
able to fully separate the contribution from $\ogg$, $\otg$, and $\ot$
while the degeneracy associated to $\optwo$ remains.\smallskip

%%%%%%%%%%%%%%%%%%%%%%%%%%%%%%%%%%%%%%%%%%%%%%%%%%%%%%%%%%%%%%%% 
%%%%%%%%%%%%%%%%%%%%%%%%%%%%%%%%%%%%%%%%%%%%%%%%%%%%%%%%%%%%%%%%

\begin{figure}[h!]
\centering
\includegraphics[width=0.45\textwidth]{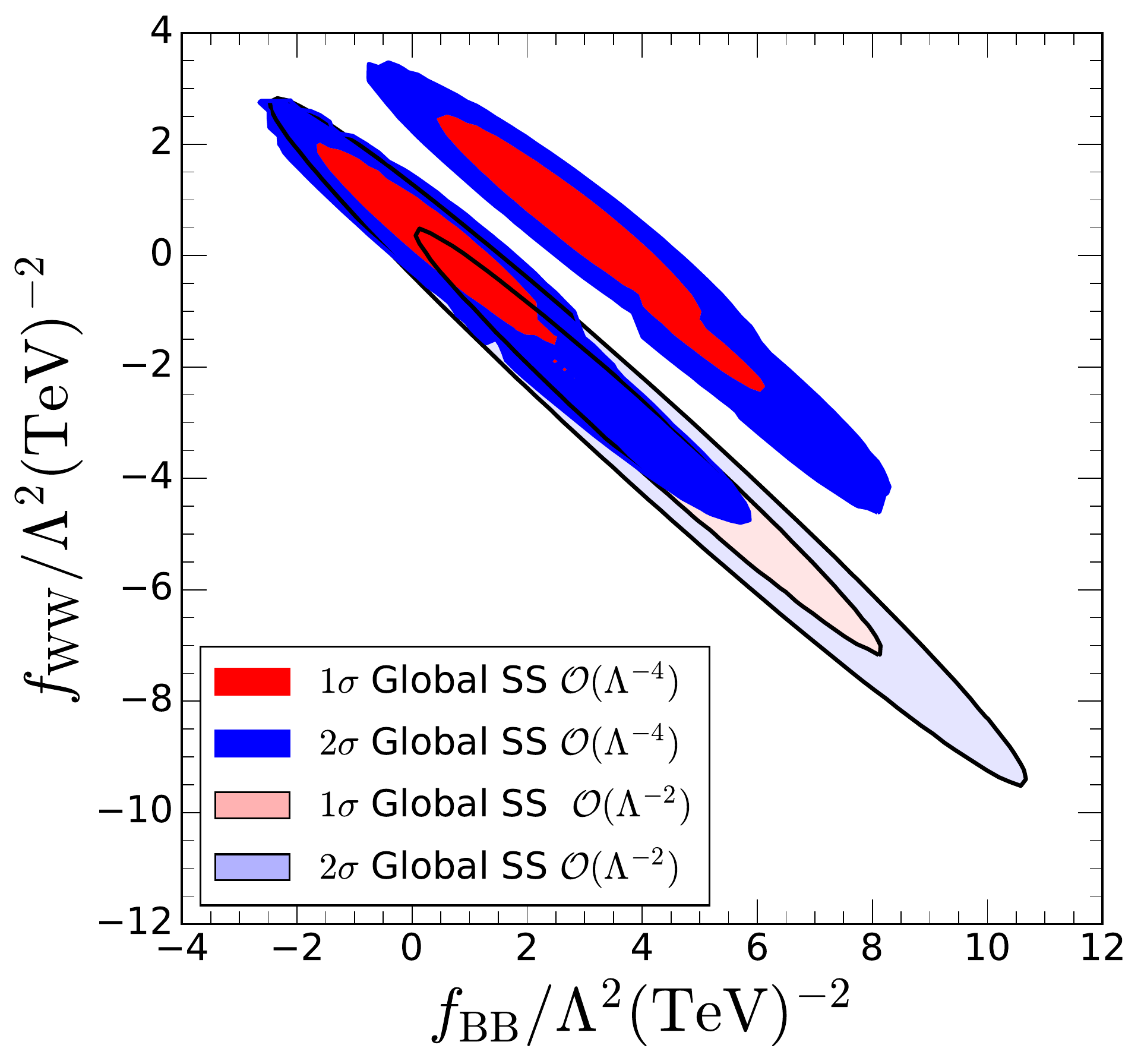}
\includegraphics[width=0.43\textwidth]{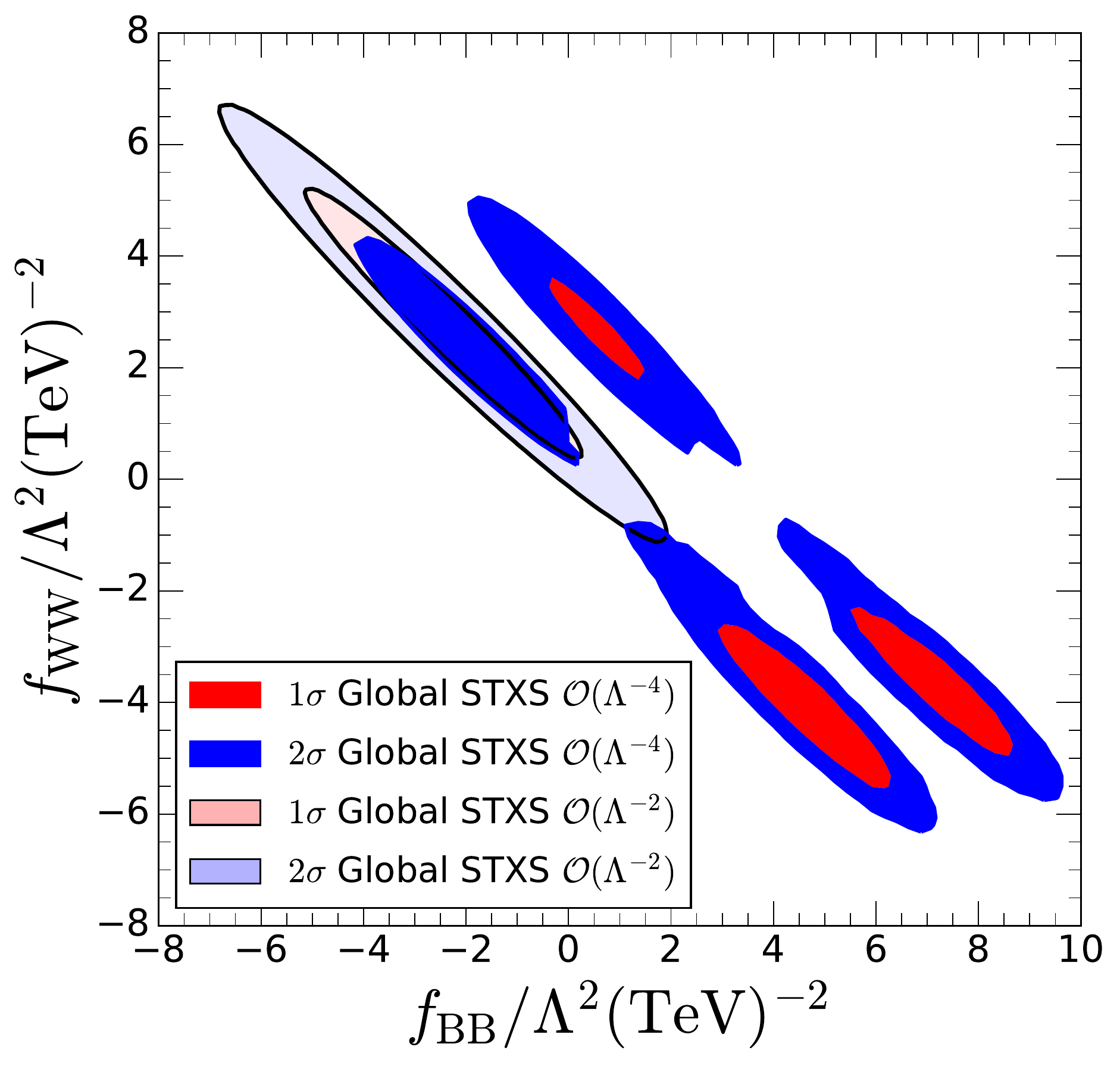}
\caption{$1\sigma$ and 95\% CL (2dof) allowed regions from the (STXS)
  SS global analysis in the plane
  $\fbb/\Lambda^2 \times \fww/\Lambda^2$ on the left (right)
  panel. These results were obtained using the theoretical
    predictions up to ${\cal O}(\Lambda^{-2})$ or up to
    ${\cal O}(\Lambda^{-4})$ approximations, as indicated by the
  color code.}
  \label{fig:corr-bbww}
\end{figure}

We now turn to the operators involved in the photon-photon-Higgs
vertex.  The upper left and central panels of Fig.~\ref{fig:bos_r2}
contain $\Delta\chi^2$ as a function of $\fbb/\Lambda^2$ and
$\fww/\Lambda^2$ respectively. The four analyses, SS and STXS with
predictions at ${\cal O}(\Lambda^{-2})$ and ${\cal O}(\Lambda^{-4})$,
lead to a unique allowed range for both Wilson couplings compatible
with the SM at $2\sigma$ level but for the STXS analysis at
${\cal O}(\Lambda^{-4})$ we still observe several minima. To better
understand these results we present in Fig.~\ref{fig:corr-bbww}
$1\sigma$ and 95\% CL (2dof) allowed regions from the SS and STXS
global analyses in the plane $\fbb/\Lambda^2 \times \fww/\Lambda^2$.
From the figures we see that all allowed regions display the strong
anti-correlation between those parameters associated with the
$(\fww+\fbb)/\Lambda^2$ dependence of the Higgs-photon-photon
vertex. This correlation is not exact because it is broken by other
measurements, in particular by the $H F_{\mu\nu} Z^{\mu\nu}$ branching
ratio which constrains a different combination $\fww/\Lambda^2$ and
$\fbb/\Lambda^2$.  In the figure we also observe the existence of the
second allowed region(s) around $(\fww+\fbb)/\Lambda^2 \sim 3$
TeV$^{-2}$ in the analysis performed to ${\cal O}(\Lambda^{-4})$,
associated to the flip of sign of the Higgs-photon-photon coupling
with respect to the SM value discussed below
Eq.\eqref{eq:vert-gaga}. In this case, unlike for the
gluon-gluon-Higgs vertex, the kinematic information contained in the
STXS observables cannot resolve these two solutions for which the
kinematics is identical.  In addition there are two solutions for each
of those mirror solutions associated to the degeneracy associated with
$\fptwo$ which affects the production cross section. For the STXS
analysis the two solutions become separated enough to lead to the two
additional disconnected regions observed in the right
panel. Further details can be obtained from the correlation
  matrices presented in appendix A.  \smallskip

%%%%%%%%%%%%%%%%%%%%%%%%%%%%%%%%%%%%%%%%%%%%%%%%%%%%%%%%%%%%%%%%

%%%%%%%%%%%%%%%%%%%%%%%%%%%%%%%%%%%%%%%%%%%%%%%%%%%%%%%%%%%%%%%%

\begin{figure}[h!]
\centering
\includegraphics[width=0.43\textwidth]{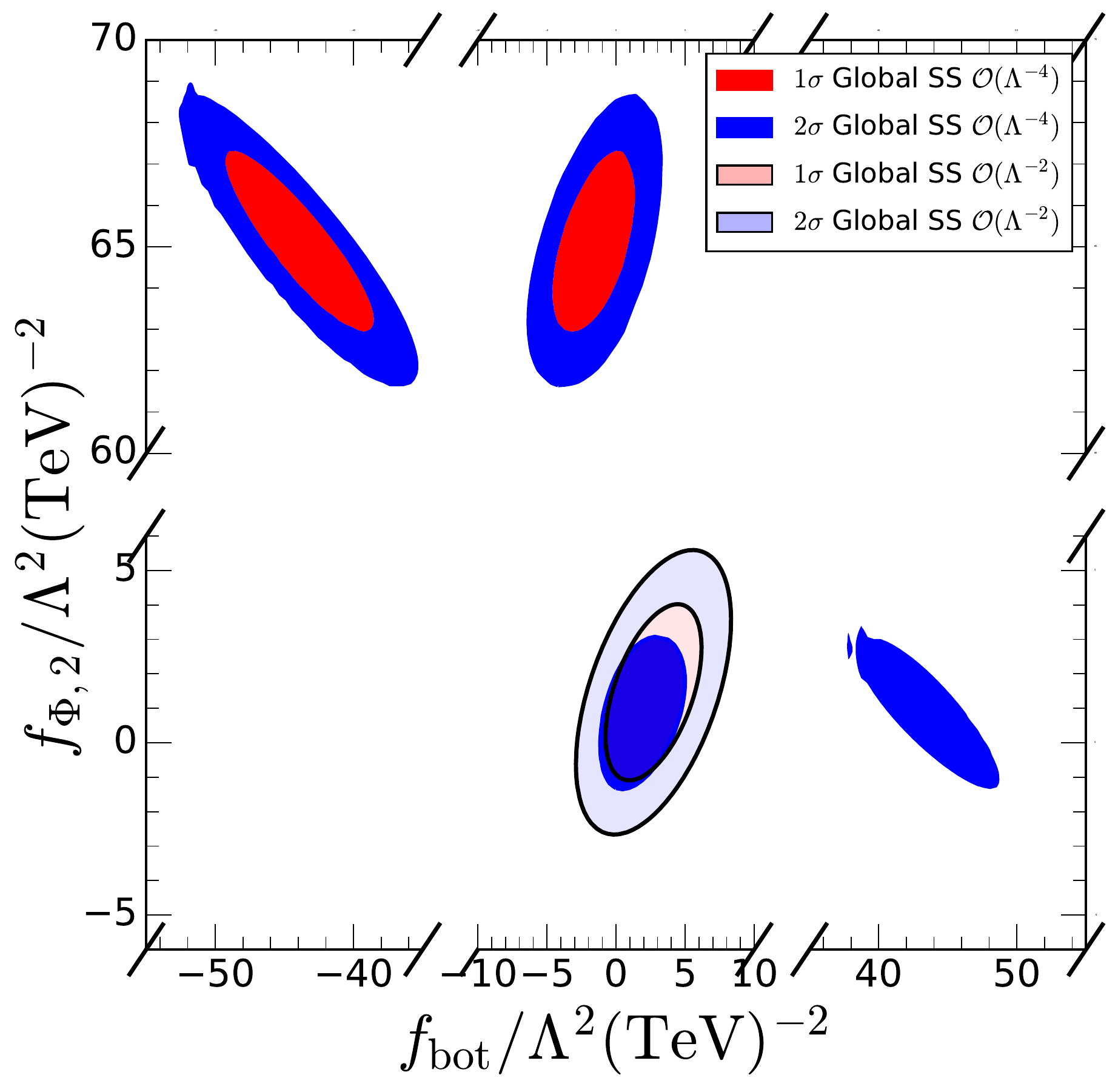}
\includegraphics[width=0.45\textwidth]{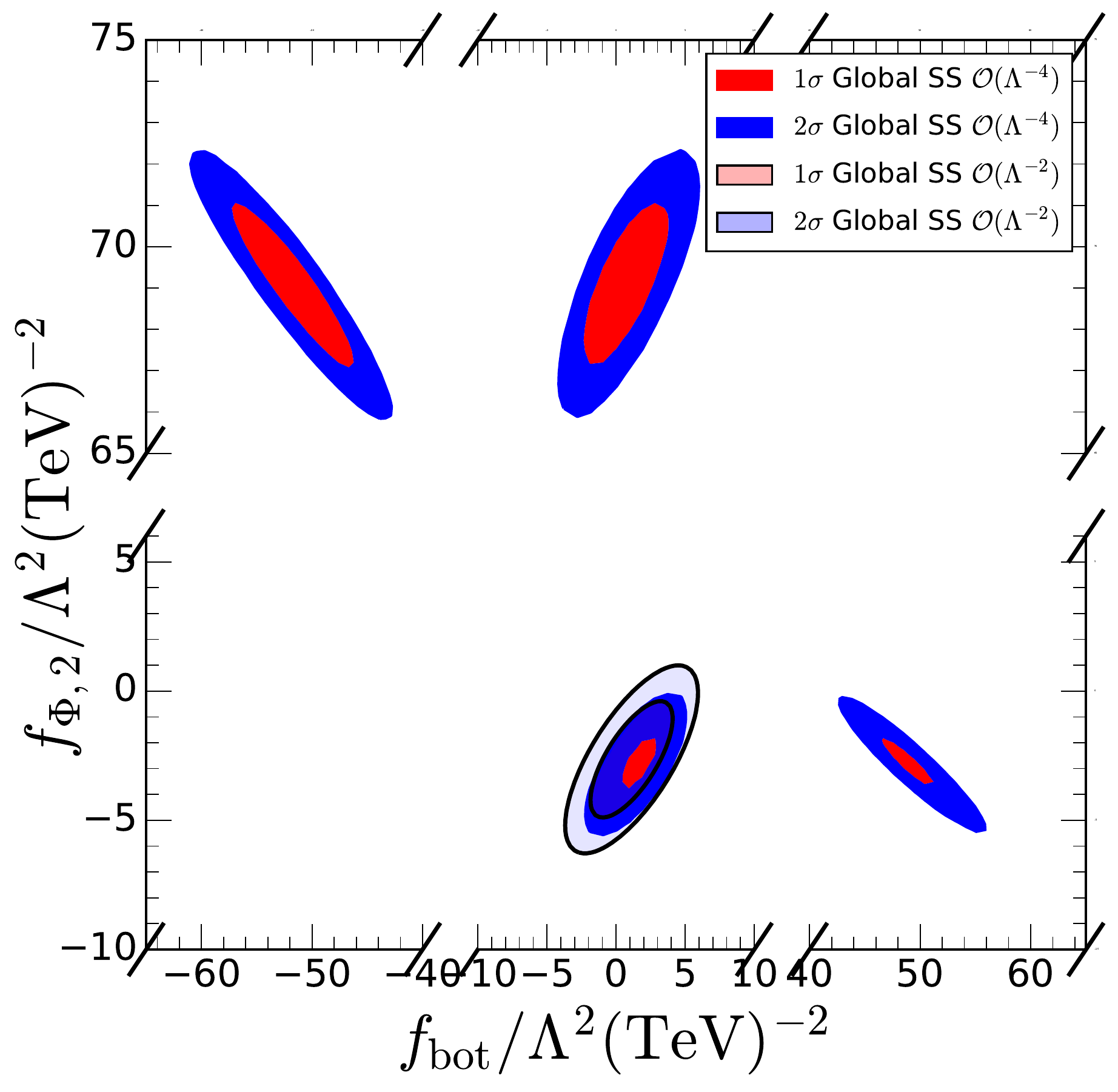}
\caption{$1\sigma$ and 95\% CL (2dof) allowed regions from the STXS
  global analysis in the plane
  $\fptwo/\Lambda^2 \times f_{b}/\Lambda^2$. These results were
  obtained using the Higgs SS and STXS data keeping the quadratic
  terms of the theoretical prediction, as indicated by the color
  code.}
  \label{fig:corr-phi2bt}
\end{figure}

In what respects the Yukawa couplings $Hff$, as discussed in
Sec.~\ref{sec:thframe}, they exhibit a four-folded degeneracy in the
${\cal O}(\Lambda^{-4})$ analysis because the sign of this coupling
can be flip by $\fptwo/\Lambda^2$ and $f_f/\Lambda^2$; see
Eq.~\eqref{eq:vert-yuk}.  This degeneracy favors the existence of
three disconnected allowed ranges~\cite{daSilvaAlmeida:2018iqo} and it
is clearly observable in both analyses at ${\cal O}(\Lambda^{-4})$ in
the three lower panels in Fig.~\ref{fig:bos_r2} for the bottom,
  $\tau$ and $\mu$ Yukawas as well as in the left panel of
  Fig.~\ref{fig:corr-phi2bt}.  These solutions are not totally
degenerate because, as discussed above, in the
${\cal O}(\Lambda^{-4})$ analysis we find a slight preference for the
non-standard solution for $\fptwo/\Lambda^2$.  Including the kinematic
information in the form of the STXS observables does not resolve this
degeneracy.  \smallskip

For $\ft/\Lambda^2$ the four-folded degeneracy is expect to be
partially broken since the scattering amplitude for the $tH$
production receives contributions from the $ttH$ and $VVH$ vertices,
therefore, being sensitive to the relative sign of the different
diagrams contributing ~\cite{Barger:2009ky,Biswas:2013xva,
  Chang:2014rfa}. Conversely, the contribution of this coupling to the
effective gluon-gluon-Higgs vertex introduces the additional
degeneracy/correlations with $\fgg$ and $\ftG$ described above.
Altogether, we find that the STXS Higgs data is able to constrain
univocally the top Yukawa coupling even at ${\cal O}(\Lambda^{-4})$.
\smallskip

%%%%%%%%%%%%%%%%%%%%%%%%%%%%%%%%%%%%%%%%%%%%%%%%%%%%%%%%%%%%%%%% 
%%%%%%%%%%%%%%%%%%%%%%%%%%%%%%%%%%%%%%%%%%%%%%%%%%%%%%%%%%%%%%%%

We present the one sigma bounds and correlations in appendix A for the
${\cal O}(\Lambda^{-2})$ global SS and STXS analysis. From these
results we can see that the strongest (anti-)correlations are between
the pairs of operators $\ob \otimes \obb$, $\ob \otimes \oww$,
$\obb \otimes \oww$, $\ogg \otimes \ot$, $\ogg \otimes \otg$,
$\obw \otimes \opone$, and $\oer \otimes \ollll$. The first three
stems mainly from the Higgs decay into photon pairs, while the next
two are due to the Higgs coupling to gluons, and the last two are
dominantly due to their contribution to the EWPD observables.
  Furthermore, $\optwo$ possesses sizable correlations with many
  dimension-six operators due to the possibility of flipping the sign
  of the Higgs couplings.   \smallskip

%%%%%%%%%%%%%%%%%%%%%%%%%%%%%%%%%%%%%%%%%%%%%%%%%%%%%%%
\section{Results for Simplified Models}
\label{sec:simpmod}

\begin{figure}[h!]
\centering
\includegraphics[width=\textwidth]{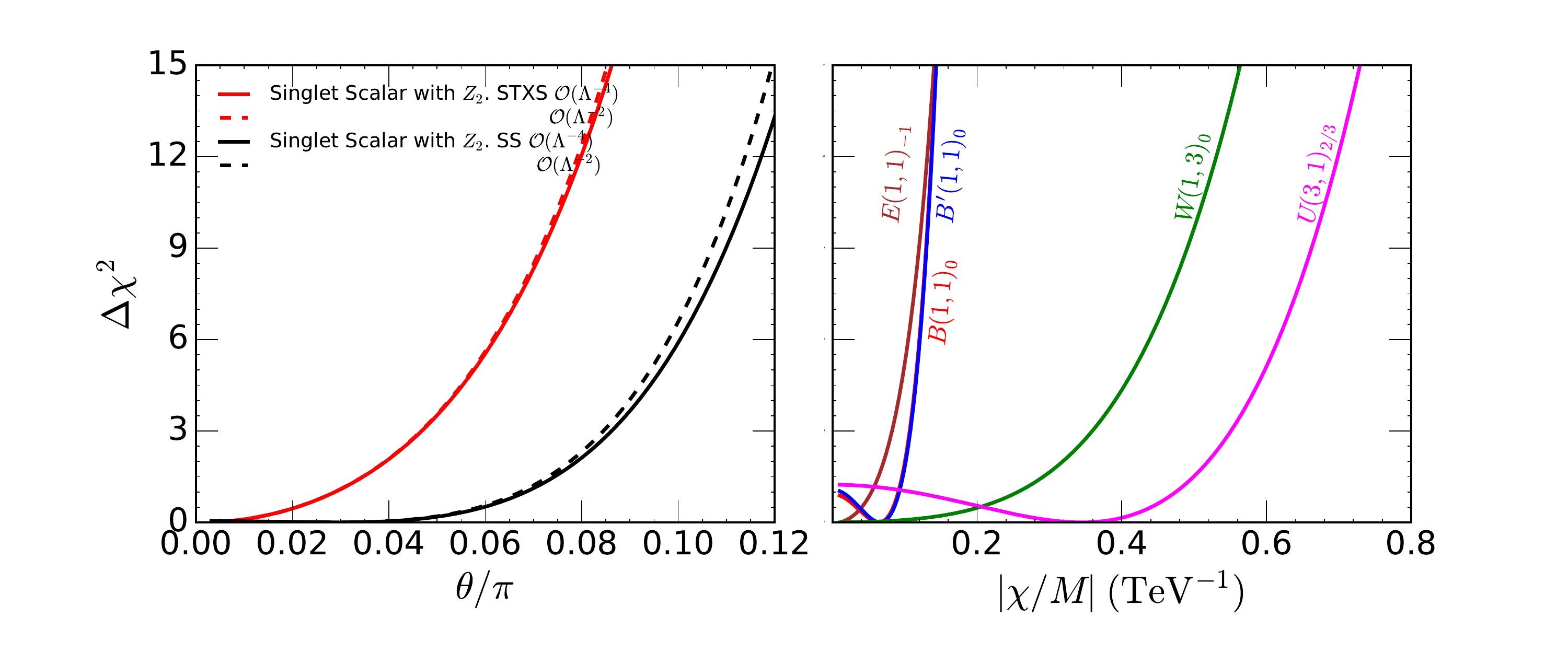}
\caption{ The left (right) panel contains $\Delta\chi^2$ as a function
  of $\theta/\pi$ ($\chi/M$) for the singlet scalar model (models with
  one additional state).  The results shown on the right panel
  correspond to the global analysis with the STXS Higgs data sets and
  including only the contributions that are linear on the Wilson
  coefficients (${\cal O}(\Lambda^{-2})$).}
 \label{fig:mod1d}
\end{figure}

\begin{figure}[h!]
\centering
\includegraphics[width=0.80\textwidth]{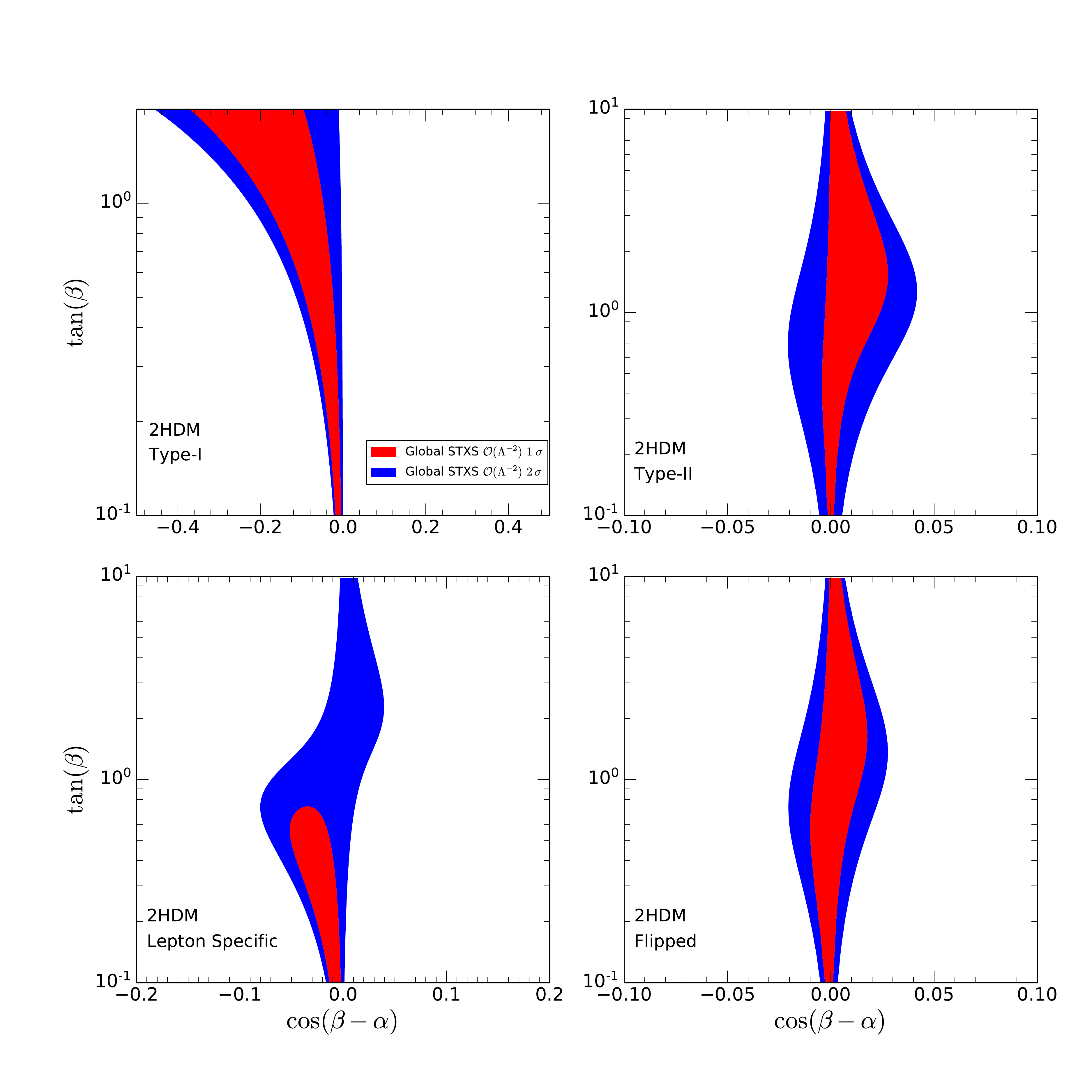}
\caption{ Allowed regions of the plane
  $\cos(\beta-\alpha) \times \tan\beta$ for several two Higgs doublet
  models. The results shown correspond to the global analyses with the
  STXS Higgs data sets including only the contributions that are
  linear on the Wilson coefficients (${\cal O}(1/\Lambda^2)$).}
 \label{fig:mod2d}
\end{figure}

Figures~\ref{fig:mod1d} and ~\ref{fig:mod2d} contain the results of
the analyses for the simplified models presented in
Sec.~\ref{sec:models}.\smallskip

The $\Delta\chi^2$ distribution for the singlet scalar extension of
the SM is presented in the left panel of Fig.~\ref{fig:mod1d}, for
both global analysis and the predictions obtained up to
${\cal O}(\Lambda^{-2}) $ and up to ${\cal O}(\Lambda^{-4})$,
which lead to similar results.  We find, for example, that
considering only the effects at order ${\cal O}(\Lambda^{-2})$ the
bound obtain from the SS analysis is
\begin{eqnarray}
  |\sin\theta| < 0.279 \;\; 
  \label{eq:boundsinglet}
\end{eqnarray}
at 95\% CL. This model only generates $\fptwo/\Lambda^2$ at tree
level, therefore, we can translate the above limit into
$\fptwo/\Lambda^2 < 1.4$ TeV$^{-2}$ using Eq.~\eqref{eq:mod-sing}.
Notice that this constraint is about 4 times stronger than the
than the limit originating from the 21-parameter analysis; see
Table~\ref{tab:ranges}.  Also, this limit is similar to that 
that derived in Ref.~\cite{ATLAS:2020qdt}.  \smallskip

%%%%%%%%%%%%%%%%%%%%%%%%%%%%%%%%%%%%%%%%%%%%%%%%%%%%%%%%%%%%%%%% 
%%%%%%%%%%%%%%%%%%%%%%%%%%%%%%%%%%%%%%%%%%%%%%%%%%%%%%%%%%%%%%%%%

The $\Delta\chi^2$ distributions for the simplified models that
contain the addition of one state described in Table~\ref{tab:models}
can be found in the right panel of Fig.~\ref{fig:mod1d}. For
concreteness, we show the results for the global STXS analysis at
${\cal O}(\Lambda^{-2})$, but for these models the inclusion of the
quadratic terms has a very little impact indicating the stability of the
results with respect to higher orders corrections as well as the
validity of the high mass expansion.  From the figure we read the
following 95\% CL bounds on $\chi^2/M$, which implies the quoted mass
limit for $\chi=1$:
\begin{eqnarray}
  \frac{\chi}{M}<0.084\;{\rm TeV}^{-1}  &\;\;\;( M>12\;{\rm TeV})\;\;\;
  & {\rm Model\;with\;}E(1,1)_{-1} \;\;,
  \nonumber\\
  \nonumber
  \frac{\chi}{M}<0.15\;{\rm TeV}^{-1}  & \;\;\;( M>6.7\;{\rm TeV})\;\;\;
  & {\rm Model\;with\;}B'(1,1)_{0} \;\;,
  \\ 
  \frac{\chi}{M}<0.16\;{\rm TeV}^{-1}  & \;\;\;( M>6.2\;{\rm TeV})\;\;\;
  & {\rm Model\;with\;}B(1,1)_{-1} \;\;,
  \label{eq:bound1par}
  \\
  \nonumber
  \frac{\chi}{M}<0.35\;{\rm TeV}^{-1}  & \;\;\;( M>2.8\;{\rm TeV})\;\;\;
  & {\rm Model\;with\;}W(1,3)_{0} \;\;,
  \\\nonumber
  \frac{\chi}{M}<0.58\;{\rm TeV}^{-1}  & \;\;\;( M>1.7\;{\rm TeV})\;\;\;
  & {\rm Model\;with\;}U(3,1)_{2/3} \;\;.
\end{eqnarray}
The tightest constraint is for the model containing a new lepton $E$
(brown curve) since this model generates the strongly bound Wilson
coefficients $\fer/\Lambda^2$ and $\fpone/\Lambda^2$, which appear
when the Wilson coefficients generated at the high scale matching
  are rotated to the HISZ basis using Eqs.~\eqref{eq:rot1}
and~\eqref{eq:rot2}.  The models with an extra vector $B$ are also
subject to strong limits because they generate $\fpone/\Lambda^2$
\emph{i.e.} they contribute to the $T$ parameter.  \smallskip

The additional vector triplet $W$ and the extra vector quark $U$
prompts the appearance of the Wilson coefficient $\fqthree/\Lambda^2$
which is well constrained by the precise determination of the
electroweak gauge couplings of left-handed quarks with  EWPD and
  LHC diboson~\cite{Alves:2018nof} and Higgs associated
  production~\cite{Biekoetter:2018ypq, Brehmer:2019gmn} data. The
extra vector quark $U$ generates also $\fqone/\Lambda^2$ but with
opposite sign. $\fqone/\Lambda^2$ also contributes to the couplings of
the left-handed quarks to the $Z$ boson, what leads to a small
anti-correlation between these two coefficients. This results into the
slighter weaker bounds in this model. \smallskip

%%%%%%%%%%%%%%%%%%%%%%%%%%%%%%%%%%%%%%%%%%%%%%%%%%%%%%%%%%%%%%%% 
%%%%%%%%%%%%%%%%%%%%%%%%%%%%%%%%%%%%%%%%%%%%%%%%%%%%%%%%%%%%%%%%%

Figure~\ref{fig:mod2d} contains the constraints on 2HDMs that we
obtain performing the global analysis at ${\cal O}(\Lambda^{-2})$ and
using the STXS Higgs data sets.  In the figure, we see that the
allowed range for $\cos(\beta-\alpha)$ is tightly constrained in
agreement with the alignment assumption with the only exception of the
Type-I model in the large $\tan\beta$ limit for which the Wilson
coefficients we are considering approach zero and, therefore, no bound
can be imposed in this approximation.  In fact, with that exception
our results show that the allowed parameter space at 95\% CL is
strongly bounded. Compared to the experimental results obtained in
Ref.~\cite{CMS:2018uag} that used just a fraction of the Run 2
integrated luminosity \footnote{ Notice also that
  Ref.~\cite{CMS:2018uag} obtained their constraints adapting the
  so-called $\kappa$-framework which contains quadratic terms in the
  anomalous couplings.} we find stronger bounds on
$\cos(\beta-\alpha)$, for fixed $\tan\beta$ by up to a factor 4
depending on the model and the sign of $\cos(\beta-\alpha)$.  Also in
our analysis we do not find the small lobe features observed in
Ref.~\cite{CMS:2018uag} for positive $\cos(\beta-\alpha)$ and large
$\tan\beta$.  Our bounds on Type-I and II models are comparable with
those derived in Ref.~\cite{Dawson:2020oco, Khosa:2021wsu}.  \smallskip

%%%%%%%%%%%%%%%%%%%%%%%%%%%%%%%%%%%%%%%%%%%%%%%%%%%%%%%%%%%%%%%%%%%%%%
\section{Discussion}
\label{sec:discussion}

In this work we have presented the results of comprehensive analyses
of low-energy electroweak precision measurements as well as LHC data
on gauge boson pair production and Higgs observables in the context of
the SMEFT. We focused on observables related to the electroweak
sector, which at present allow for precision tests of the couplings
between electroweak gauge bosons and fermions, triple electroweak
gauge couplings and the couplings of the Higgs to fermions and gauge
bosons. For the sake of assessing the impact of the Higgs kinematic
distributions we performed an analysis with and without the STXS Higgs
data in combination with the Higgs total signal strengths.  In total,
the global analyses of EWPD and EWDBD and Higgs results from LHC
encompasses 167 observables when considering only SS data and 255
observables when including the STXS samples; see Sec.~\ref{sec:frame}
for further details. \smallskip

We worked in the framework of effective lagrangians assuming the
linear realization of the electroweak gauge symmetry. Dimension-six
operators are those with lowest dimension which contribute significantly
to the considered
processes at lowest order.  The global analysis involves the 21
operators in Eq.~\eqref{eq:leff} under the flavor assumption that the
new operators do not introduce additional tree level sources of flavor
violation nor violation of universality of the weak
current. Furthermore, we also analyzed the constraints on a few
simplified models to illustrate how relations between the generated
Wilson coefficients within specific models lead to tighter
limits. \smallskip

\begin{figure}[h!]
\centering
\includegraphics[width=0.87\textwidth]{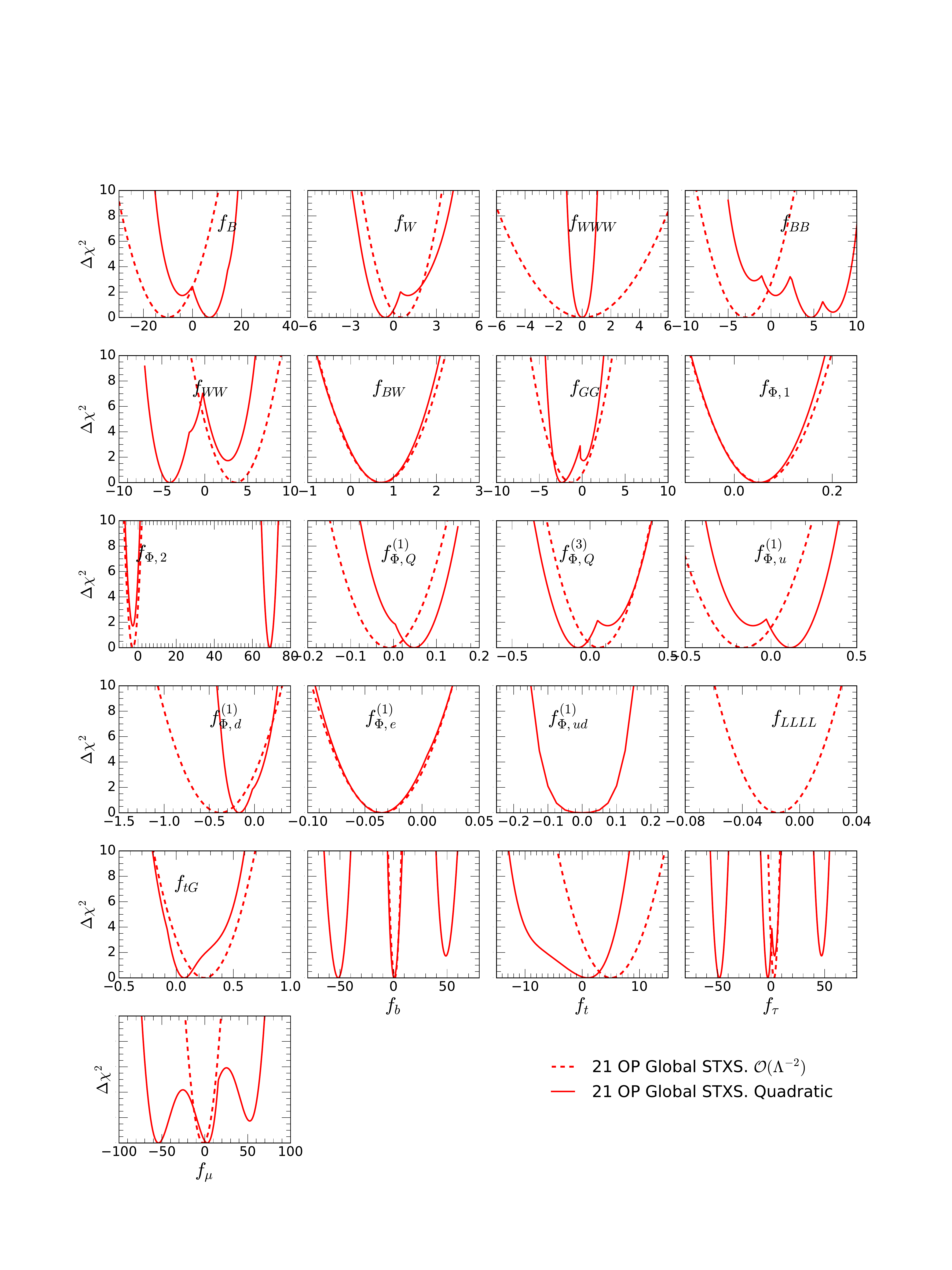}
\caption{Marginalized one-dimensional $\Delta\chi^2$ distributions for
  the 21 parameters appearing in our global fit including the STXS
  Higgs data sets. The dashed (solid) line stands for the results
  obtained with the theoretical predictions for the observables
  expanded at ${\cal O}(\Lambda^{-2})$ ( ${\cal O}(\Lambda^{-4})$) in
  the Wilson coefficients.}
 \label{fig:globstxs}
\end{figure}

\begin{figure}[h!]
\centering
\includegraphics[width=0.75\textwidth]{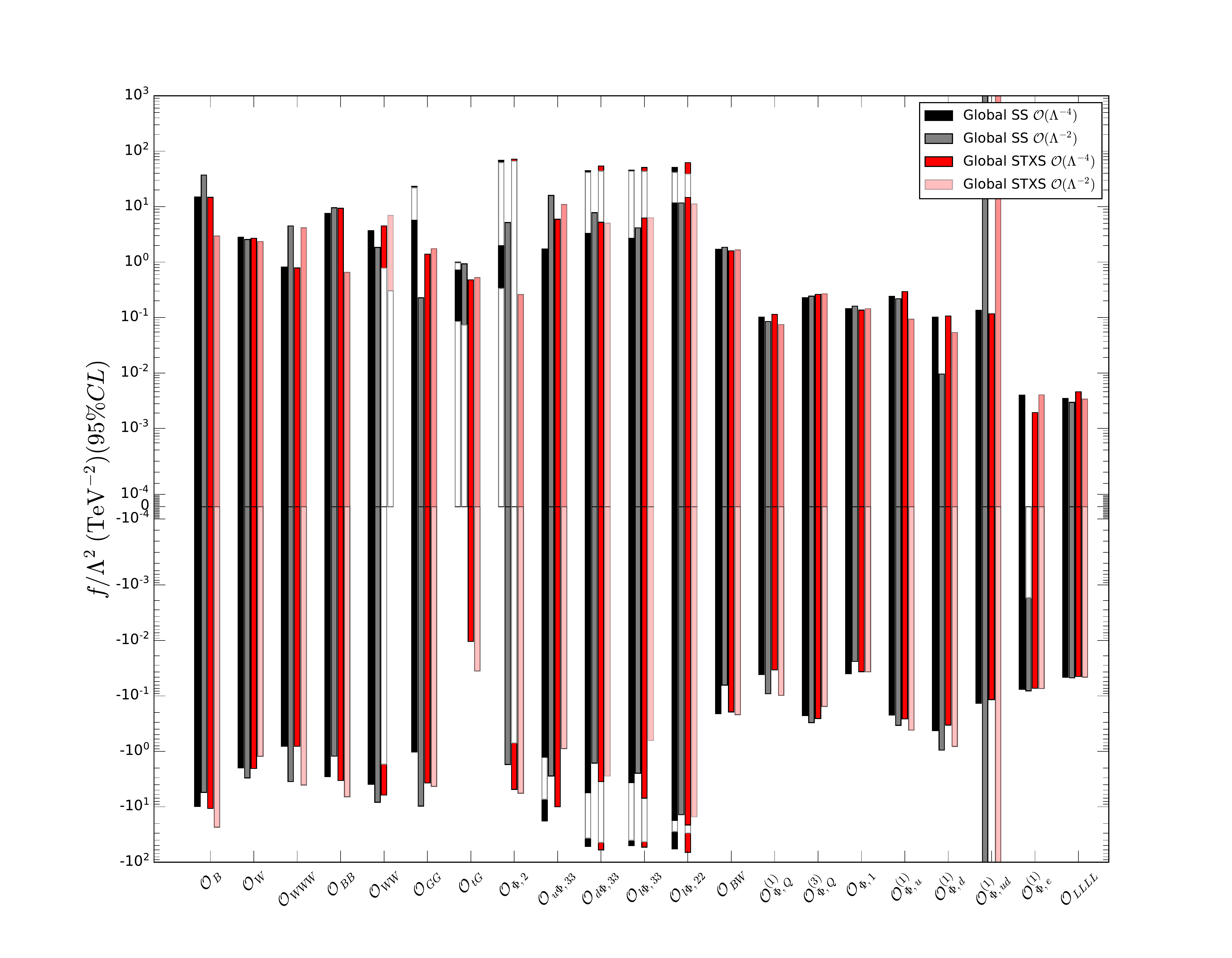}
\caption{95\% CL on the 21 Wilson coefficients used in our
  analyses. The color code indicates the data set used and the order of
  the predictions used.}
 \label{fig:franges}
\end{figure}

\begin{figure}[h!]
\centering
\includegraphics[width=0.75\textwidth]{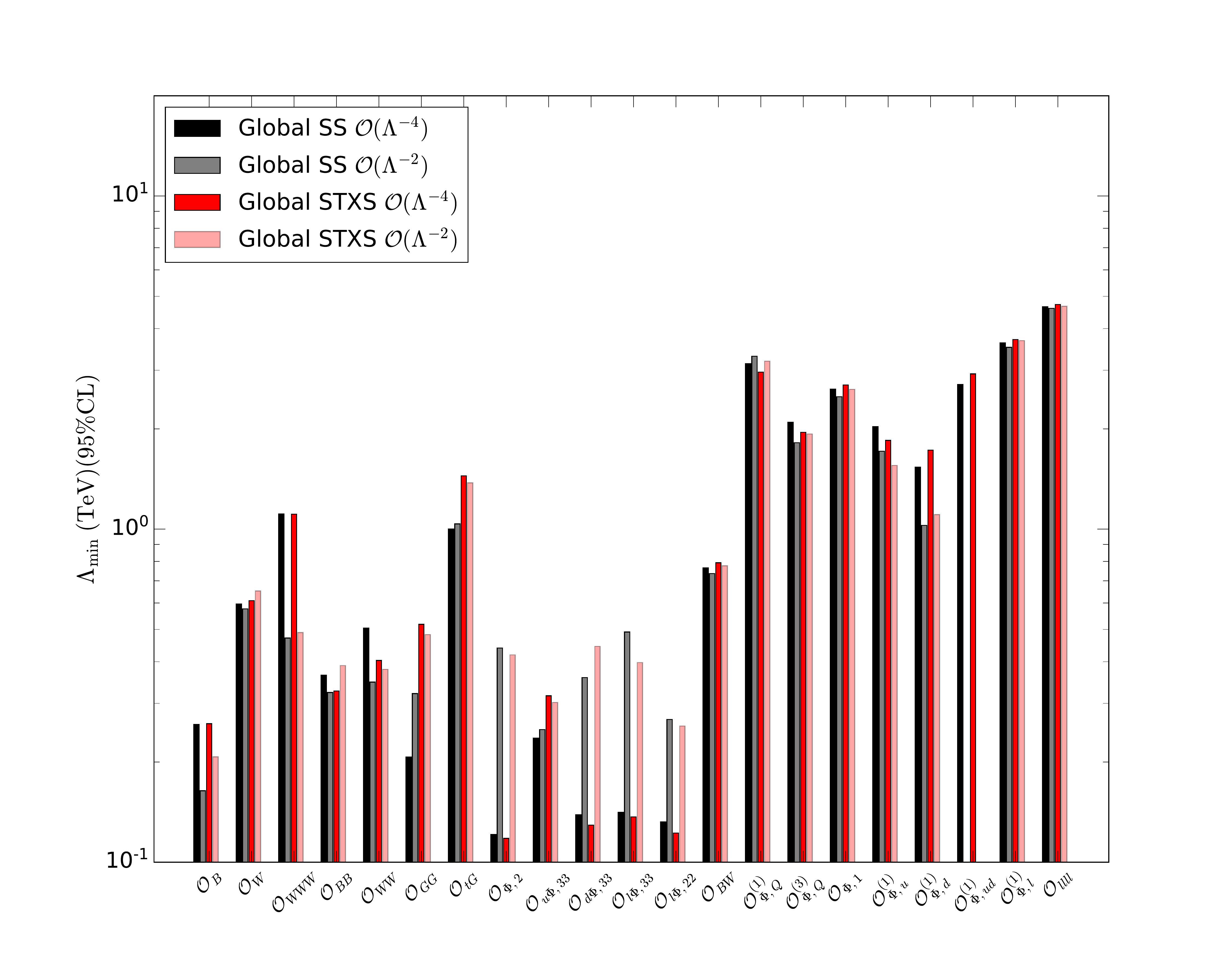}
\caption{95\% CL minimal effective scale for each of the 21 Wilson
  coefficients. The color code indicates the data set used and the order
  of the predictions used.}
 \label{fig:scales}
\end{figure}

All of the analyses performed show no statistically significant source
of tension with the SM. We find
\begin{eqnarray}
\chi^2_{\rm min\;EWPD+EWDBD,\; SM}=91\;,
&&{\rm\; 111\;observables\;,} \nonumber
\\
\chi^2_{\rm min\;Global\ SS,\; SM}=133\;,  &&
    {\rm\; 166\;observables\;,}
  \\
  \chi^2_{\rm min\;Global\ STXS,\; SM}=304\;,
  && 
{\rm 255\;observables\;},   \nonumber
\end{eqnarray}
to be compared with 
\begin{eqnarray}
\chi^2_{\rm min\;EWPD+EWDBD,\; SMEFT\,{\cal O}(\Lambda^{-4})\;[SMEFT\,{\cal O}(\Lambda^{-2})]}=87\;[85],&&
{\rm\; 111\;observables\;\&\; 12\; operators\;,}\nonumber
\\
\chi^2_{\rm min\;Global\ SS, SMEFT\,{\cal O}(\Lambda^{-4})\; [SMEFT\,{\cal O}(\Lambda^{-2})]}=115\;[112], &&
{\rm\; 166\;observables\;\& \; 21\; operators\;,} 
  \\
  \chi^2_{\rm min\;Global\ STXS,\; SMEFT\,{\cal O}(\Lambda^{-4})\;[SMEFT\,{\cal O}(\Lambda^{-2})]}=266\;[264],
&&{\rm 255\;observables\;\& \; 21\; operators\;}.
  \nonumber
\end{eqnarray}  
We summarize our results of the $\chi^2$ dependence on the Wilson
coefficients for the ${\cal O}(\Lambda^{-2})$ and
${\cal O}(\Lambda^{-4})$ analyses, which we performed with the most
comprehensive data set including the kinematic information on the
Higgs observables, by displaying the corresponding one-dimensional
$\Delta\chi^2$ distributions shown in Fig.~\ref{fig:globstxs}, where
we marginalized over the 20 undisplayed variables in each panel. With
these results and  the corresponding ones for the global SS analysis
we obtain the 95\% CL allowed ranges of the 21 Wilson coefficients that
we present in Table~\ref{tab:ranges} and graphically display them in
Fig.~\ref{fig:franges}. The maximum allowed value for each Wilson
coefficient at a given CL can be translated into a lower bound bound
on an effective new physics scale
\begin{equation}
  \Lambda_{\rm min,CL}
  \equiv \frac{1}{\sqrt{|f/\Lambda^2|_{\rm
       max,CL}}} \;\;,
\end{equation}
that are depicted in Fig.~\ref{fig:scales}. Notice that
$\Lambda_{\rm min,CL}$ only coincides with the minimum energy scale in
the operator expansion, $\Lambda$, for coefficients $f=1$. $\Lambda$
could be smaller than $\Lambda_{\rm min,CL}$ if the coupling is weak.
\smallskip

%%%%%%%%%%%%%%%%%%%%%%%%%%%%%%%%%%%%%%%%%%%%%%%%%%%%%%%
\begin{table}
\begin{tabular}{|c||c|c||c|c|}\hline
  Operator & \multicolumn{4}{c|}{95\% CL $f/\Lambda^2$ (TeV$^{-2}$)}\\
\hline
& Global SS  ${\cal O}(\Lambda^{-4})$, & Global SS  ${\cal O}(\Lambda^{-2})$,
& Global STXS   ${\cal O}(\Lambda^{-4})$, &
  Global STXS  ${\cal O}(\Lambda^{-2})$, \\\hline
  $\ob$ & (-9.8,14) &  (-5.5,37) &   (-11,15)
  & (-23,3.0) 
  \\
  $\ow$ & (-2.0,2.8)&    (-3.0,2.6) & (-2.0,2.7)
  & (-1.2,2.3)
  \\
  $\owww$ &(-0.80,0.81)& (-3.5,4.5) & (-0.81,0.78)
  & (-4.1,4.2)
  \\
  $\obb$ & (-2.8,7,5) & (-1.2,9.6) & (-3.4,9.4)
  & (-6.6,0.65)
  \\ 
  $\oww$ & (-3.9,3.7)& (-8.3,1.8) & (-6.1,-1.7) $\,\cup\,$ (0.78,4.5)
  & (0.30,7.9)
  \\
  $\ogg$ & (-1.0,5.7) $\,\cup\,$  (22,23) & (-9.7,0.23) &  (-3.7,1.4) 
  & (-4.3,1.7)
  \\
  $\otg$ & (0.11,0.71) & (0.073,0.93) & (-0.010,0.48)
  & (-0.035,-0.53)
  \\
  $\optwo$ & (0.33,2.0)$\,\cup\,$(62,68) & (-1.7,5.2)
  & (-4.7,-0.71)$\,\cup\,$(66,72)
  &(-5.7,0.26)
  \\
  $\ot$ &(-18,-7,3)$\,\cup\,$(-1.3,1.7) & (-2.8,16) & (-10,5.9)
  & (-0.89,11)
  \\
  $\obo$ & (-52,-37)$\,\cup\,$(-5.6,3.3) $\,\cup\,  $  (41,45)& (-1.6,7.8)
  & (-60,-44)$\,\cup\,  $(-3.5,5.2)$\,\cup\,  $ (44,54)    
  & (-2.8,5.0)
  \\
  $\ota$ &(-50,-40)$\,\cup\,$(-3.7,2.7) $\,\cup\,$(44,45)
  & (-2.5,4.2) 
  &(-53,-43)$\,\cup\,$(-7.0,6.2) $\,\cup\,$(43,51)
  & (-0.64,6.3)
  \\
  $\omu$ &(-57,-28)$\,\cup\,$(-18,11) $\,\cup\,$(41,51)
  & (-14,12) 
  &(-69,-30)$\,\cup\,$(-22,15) $\,\cup\,$(39,62)
  & (-15,11)
  \\
  $\obw$  & (-0.21,1.7) &(-0.064,1.8) & (-0.19,1.6)
  & (-0.22,1.7) 
  \\
  {$\opone$}
  & (-0.040,0.14) & (-0.024,0.16) & {(-0.037,0.14)}
  & {(-0.037,0.14)}
    \\   
    {$\oqthree$} & (-0.23,0.23) & (-0.30,0.24)
    & (-0.25,0.26)
    & (-0.15,0.27)  
    \\ 
    {$\oqone$} & (-0.041,0.10) & (-0.091,0.085) & {(-0.034,0.11)}
    & {(-0.098,0.075)}
    \\
    $\our$ & (-0.22,0.24) & (-0.34,0.22) & (-0.26,0.29)
    & (-0.41,0.094)
    \\
    $\odr$ & (-0.42,0.10) & (-0.95,0.0096) & (-0.34,0.11)
    & (-0.81,-0.054)
    \\
  $\oud$ & (-0.13,0.13) &  --- & (-0.12,0.12) &  ---
    \\
    $\oer$ &  (-0.076,0.0040) & (-0.081,-0.0016)& (-0.072,0.0020)
    &(-0.074,-0.0040)
    \\ 
    $\ollll$ & (-0.046,0.0035) & (-0.047,0.0029)& (-0.045,0.0046)
    & (-0.046,0.0034)
    \\\hline
\end{tabular}
\caption{Marginalized 95\% CL allowed ranges for the Wilson coefficients
  for the four different global analyses performed in this work.}
\label{tab:ranges}
\end{table}

%%%%%%%%%%%%%%%%%%%%%%%%%%%%%%%%%%%%%%%%%%%%%%%%%%%%%%%%%%%%%%%%
%%%%%%%%%%%%%%%%%%%%%%%%%%%%%%%%%%%%%%%%%%%%%%%%%%%%%%%%%%%%%%%%%

%%%%%%%%%%%%%%%%%%%%%%%%%%%%%%%%%%%%%%%%%%%%%%%%%%%%%%%%%%%%%%%%
%%%%%%%%%%%%%%%%%%%%%%%%%%%%%%%%%%%%%%%%%%%%%%%%%%%%%%%%%%%%%%%%%

%%%%%%%%%%%%%%%%%%%%%%%%%%%%%%%%%%%%%%%%%%%%%%%%%%%%%%%%%%%%%%%%
%%%%%%%%%%%%%%%%%%%%%%%%%%%%%%%%%%%%%%%%%%%%%%%%%%%%%%%%%%%%%%%%%

In brief, the main results that we would like to stress are:
\begin{itemize}
\item In our basis there is no blind direction in the electroweak
  precision observables. Therefore, the limits originating from EWPD
  are already stringent on eight Wilson coefficients. In fact, the
  EWPD dominates the bounds on $\ollll$, $\oer$, $\opone$, $\obw$, and
  $\fqone/\Lambda^2$ up to some small but not negligible contribution
  from LHC data (see Fig.~\ref{fig:fer_r2}).
\item Conversely, the global analysis results into a sizable reduction
  of the uncertainty on $\fqthree/\Lambda^2$ with respect to the
  limits obtained using only EWPD. This is expected due to the role
  played by this operator in the Higgs associated
  production~\cite{Biekoetter:2018ypq, Brehmer:2019gmn}.
\item The analysis of the EWPD of $f^{(1)}_{\Phi d}/\Lambda^2$, points
  towards a nonzero value for this coefficient due to the 2.7$\sigma$
  discrepancy between the observed $A_{\rm FB}^{0,b}$ and the SM.
  Under the assumption that the operators modifying the fermion-gauge couplings
  are generation independent, the
  inclusion of the EWDBD and the Higgs data in the analysis (either SS
  or STXS) gives rise to a result compatible with the SM at better
  than 95\% CL and with slightly reduced errors.
 In addition, the
  global constraint on $f^{(1)}_{\Phi u}/\Lambda^2$ is also slightly
  improved by combining EWPD with the LHC results.
\item Allowing for
  $f^{(1)}_{\Phi d}/\Lambda^2$ to be different for the bottom quark, 
  the combination of EWPD with the LHC observables results in the
  quoted improvement on
  the bounds for those operators contributing to the light quark
  couplings but  $f^{(1)}_{\Phi,d_{33}}$ is only marginally affected
  by the inclusion of the EWDBD and Higgs data and its best fit remains
  nonzero at $\sim 2\,\sigma$ in the global analysis.
\item The operator $\oud$ induces right-handed charged current
  couplings for quarks and it can only be bound via its quadratic
  (${\cal O}(\Lambda^{-4})$) contributions. Including those in the LHC
  observables its Wilson coefficient can be bounded with precision
  comparable to that of the other operators affecting gauge-quark
  couplings; see Fig.~\ref{fig:globstxs}.
\item The new EWDBD lead to a significantly improved sensitivity to
  $\owww$ that is the only TGC operator that does not contribute to
  the Higgs couplings. Furthermore, the constraints on $\fb/\Lambda^2$
  and $\fw/\Lambda^2$ are also significantly tightened, specially when
  combining the information from both EWDBD and Higgs data, in
  particular the STXS kinematic distributions; see Figs.~\ref{fig:tgv_r2}
  and~\ref{fig:2dfbfw}.
\item As anticipated, the recently available kinematic distributions
  of the Higgs lead not only to more stringent limits but also remove
  some the correlations between  the different operators entering
  in the gluon-gluon-Higgs interactions; see Fig.~\ref{fig:2dgluglu}.
\item Degeneracies associated with the flip of the sign of some
    Higgs couplings to gauge-bosons and Yukawa couplings, that
    originate from $\optwo$, are not resolved in the present STXS
    analysis if performed at ${\cal O}(\Lambda^{-4})$. In fact, we can
    see clearly the effect of the existence of two $\fptwo/\Lambda^2$
    local minima in many panels of Fig.~\ref{fig:globstxs}. 
\item One can check the stability of the results of the overall
  picture by comparing the bounds derived using just the $1/\Lambda^2$
  contribution with those derived with predictions containing terms up
  to $1/\Lambda^4$; see Fig.~\ref{fig:globstxs}. From the figure we
  read that the analysis performed at ${\cal O} (1/\Lambda^2)$ does
  not allow for the degenerate solutions associated with the flip of
  the sign of the Higgs couplings nor the strong correlations induced
  in the effective gluon-gluon-Higgs interaction or
  photon-photon-Higgs interaction. Generically, the bounds derived
  including $1/\Lambda^4$ terms are somewhat stronger. This is
  particularly the case for the coefficients of $\owww$ and
    ${\cal O}^{(1)}_{\Phi d}$, which dominantly contribute to
    EWDBD. This stability suggests that the power series on
    $1/\Lambda$ is under control for the range of energy probed at the
    LHC Run 2.  Notwithstanding, it is also important to notice that
    diboson production has serious theoretical problems at
    ${\cal O}(\Lambda^{-2})$ since its differential cross section is
    negative in some regions of the parameters
    space~\cite{Baglio:2017bfe} and the ${\cal O}(\Lambda^{-4})$ terms
    should be kept in that case.
%
% \item \blue{ One solution for the phase space regions where the
%      differential cross section becomes negative is to remove higher
%      energy bins. In the $W^+W^-$ diboson case it is hard to directly
%      link the reconstructed distribution with the problematic phase
%      space regions, therefore it is safer to remove the $W^+W^-$
%      dataset from the analyses. Carefully examining the lower panels
%      of Fig.~\ref{fig:tgv_r2} we can learn that the removal of this
%      dataset would loosen the limits on $f_B/\Lambda^2$ and
%      $f_W/\Lambda^2$ when considering only the EWPD and
%      EWDBD. However, the global limits should remain approximately
%      the same since they are driven by the Higgs data.  }
\item As it is well-known, EFTs have limited range of validity which
    can be signaled, for instance, by the rapid growth of the cross
    section with the energy and consequent violation of
    unitarity~\cite{Corbett:2017qgl,Corbett:2014ora}. Consequently, 
    it is a matter of concern whether the bounds on the Wilson coefficients,
    are driven by regions
    of the phase space where the EFT expansion is no longer valid.
    Unfortunately, there is no systematic way to truncate the phase space
    in the observables included in the analysis to avoid such problematic
    regions. Alternatively, the comparison of the limits obtained by
    the analyses performed at order $1/\Lambda^2$    and $1/\Lambda^4$,
    allows us to estimate the size of the higher order
    contributions which would be most relevant in those phase-space regions.
    Our results of this comparison indicate that for most operators the
    ${\cal O}$($1/\Lambda^4$) contribution is smaller than the
    leading one, indicating that, generically, the analysis
    uses EFT in its validity range.
%    An
%    alternative way to verify the EFT applicability would be to remove
%    the high energy bins from the analyses in a {\em ad hoc} way,
%    however, this is beyond the scope of our studies.  }
%
\item Contrasting with the results of our previous analysis,
  Ref.~\cite{Corbett:2015ksa,Butter:2016cvz,Alves:2018nof}, we find
  that the bounds on bosonic operators modifying the Higgs couplings
  are much more stringent once the full LHC Run 2 data is
  considered. In particular, the kinematic distributions provided in
  the STXS format allows the derivation of constraints which are a
  factor 2 to 10 stronger for the Wilson coefficients of $\ob$, $\ow$,
  $\obb$, $\ow$, and $\ogg$ at at order $1/\Lambda^2$.
\item At ${\cal O}(\Lambda^{-2})$ the precision on the Yukawa couplings
  $\ft/\Lambda^2$, $\fbo/\Lambda^2$, $\fta/\Lambda^2$, and
  $\fm/\Lambda^2$ are similar using the SS or STXS data sets and their
  allowed range is reduced by $\sim 30\%$ for $\fbo$ and $\fta$, and
  $\sim 2$ for $\fm$ with respect to our previous
  results~\cite{Alves:2018nof}.
\item Despite the added complexity to the higgs-gluon-gluon vertex
  (see Eq.~\eqref{eq:vert-gluglu} and Fig.~\ref{fig:2dgluglu}), the
  introduction of the additional contribution (${\cal O}_{tG}$) to the
  Higgs production via gluon fusion does not affect significantly the
  analysis once the independent constraint from top physics
  (Eq.~\eqref{eq:bias}) is included.  In addition, the Higgs data set is
  able to improve slightly the limits on this coefficient, favoring a
  value closer to zero, with respect to the top physics bias.
\item Finally, the study of simplified models shows that the
available data sets are able to put stringent limits on the models
parameters as we see in Figs.~\ref{fig:mod1d} and ~\ref{fig:mod2d}
and Eq.~\ref{eq:boundsinglet} and ~\ref{eq:bound1par}. 
In particular, for all 2HMD's variants our analysis results into
bounds on  $\cos(\beta-\alpha)$, for a fixed $\beta$, which are
up to a factor $\sim$ 4 stronger than those derived in 
the experimental analysis of Ref.~\cite{CMS:2018uag}.
\end{itemize}  

Altogether we find that the increased integrated luminosity gathered
at 13 TeV allows us to obtain more stringent bounds on a larger set of
anomalous interactions and to perform new tests of the SM. So far
  there is no indication of deviations from the SM predictions. \smallskip

%%%%%%%%%%%%%%%%%%%%%%%%%%%%%%%%%%%%%%%%%%%%%%%%%%%%%%%%%%%%
  \acknowledgments This work is supported in part by Conselho Nacional
  de Desenvolvimento Cent\'{\i}fico e Tecnol\'ogico (CNPq), grants
  307265/2017-0 (A.A.) and 305762/2019-2 (O.J.P.E.), and by
  Funda\c{c}\~ao de Amparo \`a Pesquisa do Estado de S\~ao Paulo
  (FAPESP) grants 2018/16921-1 and 2019/04837-9.  M.C.G-G is supported
  by spanish grant PID2019-105614GB-C21
{financed
  by MCIN/AEI/10.13039/501100011033}
  , by USA-NSF grant PHY-1915093,
  and by AGAUR (Generalitat de Catalunya) grant 2017-SGR-929. The
  authors acknowledge the support of European ITN grant
  H2020-MSCA-ITN-2019//860881-HIDDeN.
\newpage
  \appendix

  \section{Analytical expresion  of $\chi^2$
    for
    ${\cal O}(\Lambda^{-2})$ SS and  STXS analysis}
  By definition when the theoretical predictions for the observables
  considered in the analysis are expanded to linear order 
 (i.e. ${\cal O}(\Lambda^{-2})$) in the Wilson coefficients, 
  $\Delta\chi^2$ takes the form
 {  $$\Delta\chi^2=\sum_{i=1}^{N}
  \left(\frac{f_i}{\Lambda^2}-\frac{f^0_i}{\Lambda^2}\right)
\,V^{-1}_{ij}  \left(\frac{f_j}{\Lambda^2}-\frac{f^0_j}{\Lambda^2}\right)\;,$$}
 where $V$ is the covariance matrix
 $$V_{ij}\equiv\sigma_i\,\sigma_j\,\rho_{ij}\;.$$
 For the SS analysis we find the best fit values and uncertainties\\[+0.3cm]
 
 %\begin{scalebox}{0.8}{
\resizebox{\textwidth}{!} 
 {$
 \begin{array}{|c|cccccccccccccccccccc|}
   \hline
&    \ob &\ow & \owww & \obb& \oww & \ogg& \otg& \optwo
    & \ot& \obo & \ota & \omu & \obw & \opone
 & \oqone& \oqthree & \our & \odr & \oer& \ollll\\[+0.2cm]\hline
    \frac{f^0}{\Lambda^2} ({\rm TeV})^{-2} &
                     16 &  -0.23 &  0.49 &  4.2 &  -3.2 &  -4.7 &   0.52 &   1.7 &   6.6 &   3.1 &  0.83 &  -1.1 &   0.89 & 0.068& {-0.003}&{ -0.030} &  -0.061 &  -0.47 &  -0.041 &  -0.022\\[+0.2cm]\hline
    { \sigma }&  11 &   1.4 &   2.0 &   2.7 &   2.5 &   2.5 &  -0.21 &   1.7 &   4.7 &   2.3 &   1.7 &   6.4 &   0.48 &  0.046 & {0.044}&  {0.14} &  0.14 &   0.24 &   0.020 &   0.013\\
\hline
 \end{array}$}\\[+0.3cm]

 \noindent with correlation matrix in the same order of operators as
 the above table is
  \\[+0.3cm]

 \resizebox{\textwidth}{!}
 {$\;\;\;\;\rho=\left(\begin{array}{cccccccccccccccccccc}
1.000 & 0.345 & -.065 & 0.945 & -.965 & -.045 & -.025 & 0.575 & -.155 & 0.585 & 0.185 & 0.035 & 0.135 & 0.105 & 0.025 & 0.095 & -.135 & 0.075 & -.285 & -.245\\
0.345 & 1.000 & -.185 & 0.265 & -.335 & -.045 & -.035 & -.055 & -.085 & 0.035 & 0.175 & 0.005 & 0.125 & 0.145 & -.135 & 0.755 & -.455 & 0.455 & -.295 & -.285\\
-.065 & -.185 & 1.000 & -.045 & -.015 & -.035 & -.095 & -.015 & -.045 & -.015 & -.075 & -.055 & -.095 & -.025 & -.045 & 0.035 & -.095 & 0.005 & -.075 & -.105\\
0.945 & 0.265 & -.045 & 1.000 & -.995 & -.095 & -.035 & 0.605 & -.165 & 0.615 & 0.145 & 0.035 & 0.235 & 0.245 & 0.055 & 0.065 & 0.045 & 0.045 & -.315 & -.345\\
-.965 & -.335 & -.015 & -.995 & 1.000 & -.055 & -.065 & -.695 & 0.055 & -.705 & -.215 & -.125 & -.145 & -.165 & -.045 & -.125 & -.015 & -.095 & 0.095 & 0.045\\
-.045 & -.045 & -.035 & -.095 & -.055 & 1.000 & -.955 & -.075 & -.945 & -.015 & 0.055 & -.055 & -.045 & -.045 & -.055 & 0.025 & -.055 & -.045 & -.035 & -.045\\
-.025 & -.035 & -.095 & -.035 & -.065 & -.955 & 1.000 & -.035 & 0.725 & -.015 & -.035 & -.035 & -.045 & -.035 & -.045 & -.045 & -.035 & -.055 & -.045 & -.055\\
0.575 & -.055 & -.015 & 0.605 & -.695 & -.075 & -.035 & 1.000 & -.315 & 0.485 & -.115 & -.015 & -.055 & 0.005 & -.115 & -.045 & 0.005 & 0.035 & -.135 & -.125\\
-.155 & -.085 & -.045 & -.165 & 0.055 & -.945 & 0.725 & -.315 & 1.000 & -.025 & -.035 & -.045 & -.075 & -.065 & -.045 & -.065 & -.055 & -.045 & -.025 & -.025\\
0.585 & 0.035 & -.015 & 0.615 & -.705 & -.015 & -.015 & 0.485 & -.025 & 1.000 & 0.305 & 0.025 & 0.025 & -.025 & 0.025 & -.135 & 0.015 & -.035 & -.115 & -.135\\
0.185 & 0.175 & -.075 & 0.145 & -.215 & 0.055 & -.035 & -.115 & -.035 & 0.305 & 1.000 & 0.005 & -.005 & -.005 & -.055 & 0.095 & -.115 & 0.045 & -.095 & -.085\\
0.035 & 0.005 & -.055 & 0.035 & -.125 & -.055 & -.035 & -.015 & -.045 & 0.025 & 0.005 & 1.000 & -.035 & -.035 & -.055 & -.015 & -.055 & -.025 & -.055 & -.065\\
0.135 & 0.125 & -.095 & 0.235 & -.145 & -.045 & -.045 & -.055 & -.075 & 0.025 & -.005 & -.035 & 1.000 & 0.865 & 0.045 & 0.115 & 0.165 & -.425 & -.955 & -.885\\
0.105 & 0.145 & -.025 & 0.245 & -.165 & -.045 & -.035 & 0.005 & -.065 & -.025 & -.005 & -.035 & 0.865 & 1.000 & 0.035 & 0.095 & 0.285 & -.435 & -.955 & -.715\\
0.025 & -.135 & -.045 & 0.055 & -.045 & -.055 & -.045 & -.115 & -.045 & 0.025 & -.055 & -.055 & 0.045 & 0.035 & 1.000 & -.125 & 0.505 & 0.615 & -.175 & -.205\\
0.095 & 0.755 & 0.035 & 0.065 & -.125 & 0.025 & -.045 & -.045 & -.065 & -.135 & 0.095 & -.015 & 0.115 & 0.095 & -.125 & 1.000 & -.665 & 0.695 & -.215 & -.215\\
-.135 & -.455 & -.095 & 0.045 & -.015 & -.055 & -.035 & 0.005 & -.055 & 0.015 & -.115 & -.055 & 0.165 & 0.285 & 0.505 & -.665 & 1.000 & -.045 & -.425 & -.255\\
0.075 & 0.455 & 0.005 & 0.045 & -.095 & -.045 & -.055 & 0.035 & -.045 & -.035 & 0.045 & -.025 & -.425 & -.435 & 0.615 & 0.695 & -.045 & 1.000 & 0.365 & 0.195\\
-.285 & -.295 & -.075 & -.315 & 0.095 & -.035 & -.045 & -.135 & -.025 & -.115 & -.095 & -.055 & -.955 & -.955 & -.175 & -.215 & -.425 & 0.365 & 1.000 & 0.725\\
-.245 & -.285 & -.105 & -.345 & 0.045 & -.045 & -.055 & -.125 & -.025 & -.135 & -.085 & -.065 & -.885 & -.715 & -.205 & -.215 & -.255 & 0.195 & 0.725 & 1.000
   \end{array}\right)\;.$}
\newpage
 For the STXS analysis we find the best fit values and uncertainties\\[+0.3cm]
 
 %\begin{scalebox}{0.8}{
\resizebox{\textwidth}{!} 
 {$
 \begin{array}{|c|cccccccccccccccccccc|}
   \hline
&    \ob &\ow & \owww & \obb& \oww & \ogg& \otg& \optwo
    & \ot& \obo & \ota & \omu & \obw & \opone
 & \oqone& \oqthree & \our & \odr & \oer& \ollll\\[+0.2cm]\hline
    \frac{f^0}{\Lambda^2} ({\rm TeV})^{-2} &
             -10. &   0.56 & 0.061 &  -2.9 &   3.6 &  -1.3 &   0.25 &  -2.7 &   5.0 &   1.1 &   2.8 &  -2.0 &   0.72 & {0.054}&  {-0.012} &   {0.057} &  -0.16 &  -0.38 &  -0.035 &  -0.021\\[+0.2cm]\hline
{\sigma}& 6.6 &   0.90 &   2.1 &   1.8 &  -1.7 &   1.5 &   0.14 &   1.5 &   3.0 &   2.0 &   1.7 &   6.6 &   0.47 & {0.045}&   {0.043} &   {0.11}  &   0.13 &   0.22 &   0.019 &   0.012    \\
\hline
 \end{array}$}\\[+0.3cm]

 \noindent with correlation matrix in the same order of operators as
 the above table is
  \\[+0.3cm]

 \resizebox{\textwidth}{!}
 {$\;\;\;\;\rho=\left(\begin{array}{cccccccccccccccccccc}
1.000 & -.305 & -.035 & 0.855 & -.935 & -.215 & 0.095 & 0.365 & -.095 & 0.345 & 0.085 & 0.005 & 0.255 & 0.205 & 0.035 & -.335 & 0.285 & -.235 & -.255 & -.315\\
-.305 & 1.000 & -.115 & -.375 & 0.325 & -.155 & -.045 & -.555 & 0.225 & -.435 & -.005 & -.055 & 0.155 & 0.135 & -.235 & 0.705 & -.295 & 0.305 & -.295 & -.255\\
-.035 & -.115 & 1.000 & -.035 & -.035 & -.035 & -.065 & 0.015 & -.065 & 0.025 & -.075 & -.045 & -.055 & -.045 & -.105 & 0.115 & -.125 & 0.025 & 0.015 & -.015\\
0.855 & -.375 & -.035 & 1.000 & -.975 & -.155 & 0.085 & 0.465 & -.215 & 0.405 & 0.065 & 0.005 & 0.365 & 0.315 & 0.025 & -.445 & 0.355 & -.325 & -.385 & -.275\\
-.935 & 0.325 & -.035 & -.975 & 1.000 & 0.095 & -.135 & -.535 & 0.235 & -.495 & -.125 & -.085 & -.185 & -.165 & -.135 & 0.365 & -.405 & 0.215 & -.015 & -.005\\
-.215 & -.155 & -.035 & -.155 & 0.095 & 1.000 & -.875 & -.085 & -.845 & -.075 & -.025 & -.065 & -.175 & -.185 & -.045 & -.065 & -.065 & -.055 & 0.135 & 0.055\\
0.095 & -.045 & -.065 & 0.085 & -.135 & -.875 & 1.000 & 0.145 & 0.495 & 0.155 & 0.005 & 0.015 & 0.005 & 0.055 & 0.025 & -.075 & 0.005 & -.095 & -.185 & -.125\\
0.365 & -.555 & 0.015 & 0.465 & -.535 & -.085 & 0.145 & 1.000 & -.285 & 0.625 & -.005 & -.035 & -.075 & -.045 & -.035 & -.465 & 0.305 & -.295 & -.125 & -.035\\
-.095 & 0.225 & -.065 & -.215 & 0.235 & -.845 & 0.495 & -.285 & 1.000 & -.055 & -.045 & -.055 & 0.045 & 0.045 & -.035 & 0.165 & -.165 & 0.035 & -.145 & -.115\\
0.345 & -.435 & 0.025 & 0.405 & -.495 & -.075 & 0.155 & 0.625 & -.055 & 1.000 & 0.105 & -.025 & 0.115 & -.005 & 0.035 & -.345 & 0.165 & -.195 & -.095 & -.035\\
0.085 & -.005 & -.075 & 0.065 & -.125 & -.025 & 0.005 & -.005 & -.045 & 0.105 & 1.000 & -.025 & -.095 & -.035 & -.065 & -.075 & -.015 & -.075 & -.065 & -.045\\
0.005 & -.055 & -.045 & 0.005 & -.085 & -.065 & 0.015 & -.035 & -.055 & -.025 & -.025 & 1.000 & -.035 & -.035 & -.045 & -.055 & -.035 & -.055 & -.045 & -.045\\
0.255 & 0.155 & -.055 & 0.365 & -.185 & -.175 & 0.005 & -.075 & 0.045 & 0.115 & -.095 & -.035 & 1.000 & 0.865 & 0.045 & 0.035 & 0.325 & -.325 & -.955 & -.885\\
0.205 & 0.135 & -.045 & 0.315 & -.165 & -.185 & 0.055 & -.045 & 0.045 & -.005 & -.035 & -.035 & 0.865 & 1.000 & 0.045 & -.015 & 0.405 & -.355 & -.935 & -.745\\
0.035 & -.235 & -.105 & 0.025 & -.135 & -.045 & 0.025 & -.035 & -.035 & 0.035 & -.065 & -.045 & 0.045 & 0.045 & 1.000 & -.395 & 0.575 & 0.475 & -.215 & -.235\\
-.335 & 0.705 & 0.115 & -.445 & 0.365 & -.065 & -.075 & -.465 & 0.165 & -.345 & -.075 & -.055 & 0.035 & -.015 & -.395 & 1.000 & -.545 & 0.335 & -.085 & -.055\\
0.285 & -.295 & -.125 & 0.355 & -.405 & -.065 & 0.005 & 0.305 & -.165 & 0.165 & -.015 & -.035 & 0.325 & 0.405 & 0.575 & -.545 & 1.000 & 0.285 & -.515 & -.405\\
-.235 & 0.305 & 0.025 & -.325 & 0.215 & -.055 & -.095 & -.295 & 0.035 & -.195 & -.075 & -.055 & -.325 & -.355 & 0.475 & 0.335 & 0.285 & 1.000 & 0.195 & 0.175\\
-.255 & -.295 & 0.015 & -.385 & -.015 & 0.135 & -.185 & -.125 & -.145 & -.095 & -.065 & -.045 & -.955 & -.935 & -.215 & -.085 & -.515 & 0.195 & 1.000 & 0.745\\
-.315 & -.255 & -.015 & -.275 & -.005 & 0.055 & -.125 & -.035 & -.115 & -.035 & -.045 & -.045 & -.885 & -.745 & -.235 & -.055 & -.405 & 0.175 & 0.745 & 1.000
\end{array}\right)\;.$} 
\bigskip

%%%%%%%%%%%%%%%%%%%%%%%%%%%%%%%%%%%%%%%%%%%%%%%%%%%%%%%%%%%%%%%%%%%%%%
\bibliography{references}
\end{document}